\documentclass[twocolumn,english,prl,amssymb,aps,superscriptaddress,showpacs,twocolumn,amsmath,showkeys,floatfix]{revtex4-1}
\usepackage[T1]{fontenc}
\usepackage[latin9]{inputenc}
\usepackage[active]{srcltx}
\usepackage{textcomp}
\usepackage{amsmath}
\usepackage{graphicx}
\usepackage{color}
\usepackage{esint}
\usepackage{caption}
\usepackage{physics}
\usepackage{subcaption}

\makeatletter
\usepackage{babel}

\makeatother

\begin{document}

\title {Generalised expression of the noise figure of phase sensitive amplifiers for an arbitrary number of modes}
\author{Yousra Bouasria}
\affiliation{Equipe Sciences de la Mati\`ere et du Rayonnement (ESMaR), Facult\'e des Sciences, Universit\'e Mohammed V, Rabat, Morocco}
\author{Debanuj Chatterjee}
\affiliation{Universit\'e Paris-Saclay, CNRS, ENS Paris-Saclay, CentraleSup\'elec, LuMIn, Gif-sur-Yvette, France}
\author{Fabienne Goldfarb}
\affiliation{Universit\'e Paris-Saclay, CNRS, ENS Paris-Saclay, CentraleSup\'elec, LuMIn, Gif-sur-Yvette, France}
\author{Yassine Hassouni}
\affiliation{Equipe Sciences de la Mati\`ere et du Rayonnement (ESMaR), Facult\'e des Sciences, Universit\'e Mohammed V, Rabat, Morocco}
\author{Fabien Bretenaker}
\affiliation{Universit\'e Paris-Saclay, CNRS, ENS Paris-Saclay, CentraleSup\'elec, LuMIn, Gif-sur-Yvette, France}
\affiliation{Light and Matter Physics Group, Raman Research Institute, Bangalore 560080, India}

\begin{abstract} 
Phase sensitive amplifiers (PSA), contrary to usual phase insensitive amplifiers (PIA), are in principle capable to achieve noiseless amplification, i.e. exhibit a quantum-limited noise figure (NF) of 0 dB. When implemented using four-wave mixing (FWM) in a nonlinear fiber, extra waves can be generated by undesired FWM processes, 
which may introduce extra input ports for vacuum fluctuations, thus potentially degrading the NF. In this situation, we give here a general analytical quantum derivation of the PSA NF, valid for an arbitrary number of nonlinearly coupled modes. This expression is usable as soon as a linear input-output relation can be found for the annihilation and creation operators of the involved modes. It predicts that the noise level depends on the number of interacting waves. We illustrate the usefulness of this expression in the case of six waves, corresponding to four interacting quantum modes. In this example the signal NF is degraded by 0.4 dB, compared to 10 dB obtained for PIA operation of the same scheme.
\end{abstract}


\maketitle
\section{Introduction}
The evolution of today's fibre optical communication systems strongly depends on the noise performances of the optical amplifiers and on the fibre nonlinearities. Conventional phase-insensitive amplifiers (PIAs), whose gain is independent of the signal phase, exhibit a minimum quantum-limited noise figure (NF) of 3 dB at high gain, meaning that the signal-to-noise ratio (SNR) is degraded by at least 3 dB. Phase-sensitive amplifiers (PSAs), whose gain depends on the signal phase, have a quantum limited NF of 0 dB, offering therefore the possibility of noiseless amplification of the signal, i.e. without degradation of the SNR \cite{Caves, Levenson}.  Thanks to this unique phase sensitive property and the associated noiseless amplification capability, transmission systems based on PSAs have gained a considerable attention. This have given rise to a large variety of potential applications such as generation of squeezed states \cite{Walls, Slusher, Teich}, low-noise amplification \cite{Yamamoto, Tong}, phase and amplitude regeneration \cite{Umeki}, wavelength conversion \cite{Inoue1994} and all-optical signal processing \cite{Radic2005}. 

In recent years, various research groups have focused on the quality of optical amplifiers in terms of noise, and several in-depth theoretical studies of the noise figure for both PIA and PSA systems have been proposed  \cite{C.J.McKinstrie,Marhic,Ferrini,Inoue}. Most of these investigations rely on the 3-wave or 4-wave models, leading with pump non-depletion approximation to the so-called two-mode model (TM). However, in practice, even if only 3 or 4 waves are launched at the input of the optical fibre, extra waves may be generated due to the large nonlinearity of the fibre. Intuitively, as a parametric amplifier generates extra waves, it also introduces extra noise during the amplification process. Therefore,  the analysis of the noise performances taking the existence of high-order waves into account is needed. To this aim, in Ref.\,\cite{Colin}, \emph{McKinstrie \& al.} have studied analytically the noise figure in the framework of a 6-wave model. With extra approximations (no pump depletion and negligible dispersion) this led them to the so-called four-mode model (FM) from which they could derive a generalised expression of the NF for a PIA system in many-mode configuration. More recently, in Ref.\,\cite{inoue2019influence}, \emph{Inoue}  developed a semi-analytical approach to calculate the noise figure in the framework of 7-wave model for a dual-pump PSA system. His approach describes the signal evolution quantum-mechanically, while the classical pump depletion is treated numerically. However, both of these previous works do not provide a complete theoretical description for the NF in PSA configuration when multiple waves are accounted for.  Investigating the NF in the case of many-mode model is a major challenge for two main reasons: First, as the many nonlinear interaction processes between the propagating waves occur simultaneously along the fibre, it is difficult to end up with an analytical solution for the nonlinear coupled equations, and most of the attempts made during the past few years relied on some simplifying assumptions. Second, as the number of  interacting waves taken into account increases, the expression of the NF becomes difficult to calculate, compared to the two-mode model.

Consequently, the present paper aims at addressing the second reason and proposes a generalised expression that facilitates the calculation of the NF in the case of many-mode PSA systems, while assuming that the amplifier is operating in the linear regime where the input modes are related to the output modes by a transfer matrix, i.e., an input-output linear relation. This work is thus an extension to the study tackled in Ref.\,\cite{Colin}, in which a general form of the NF in many-mode PIA systems was derived. 

The paper is structured as follows: \textcolor{black}{Section \ref{sec.1N1} reminds the basic features of PSAs based on nonlinear fibers. Then, in} Section \ref{sec.2}, the classical and quantum equations that describe the propagation of the field mode in fibres are recalled, and the assumptions used to derive an input-output linear relation for some particular models are discussed. In Section \ref{sec.3}, the noise properties of both two-mode and three-mode models are explicitly analysed by means of the quantum approach, and their results are used to deduce a generalised expression of the noise figure for models involving a large number of modes. In Section \ref{sec.4}, the generalised NF expression introduced in Section \ref{sec.3} is applied to the four-mode model, and the results are discussed for both PIA and PSA configurations. A summary is given in Section \ref{sec.5}.

\begin{figure}[h]
\centering
\includegraphics[width=\columnwidth]{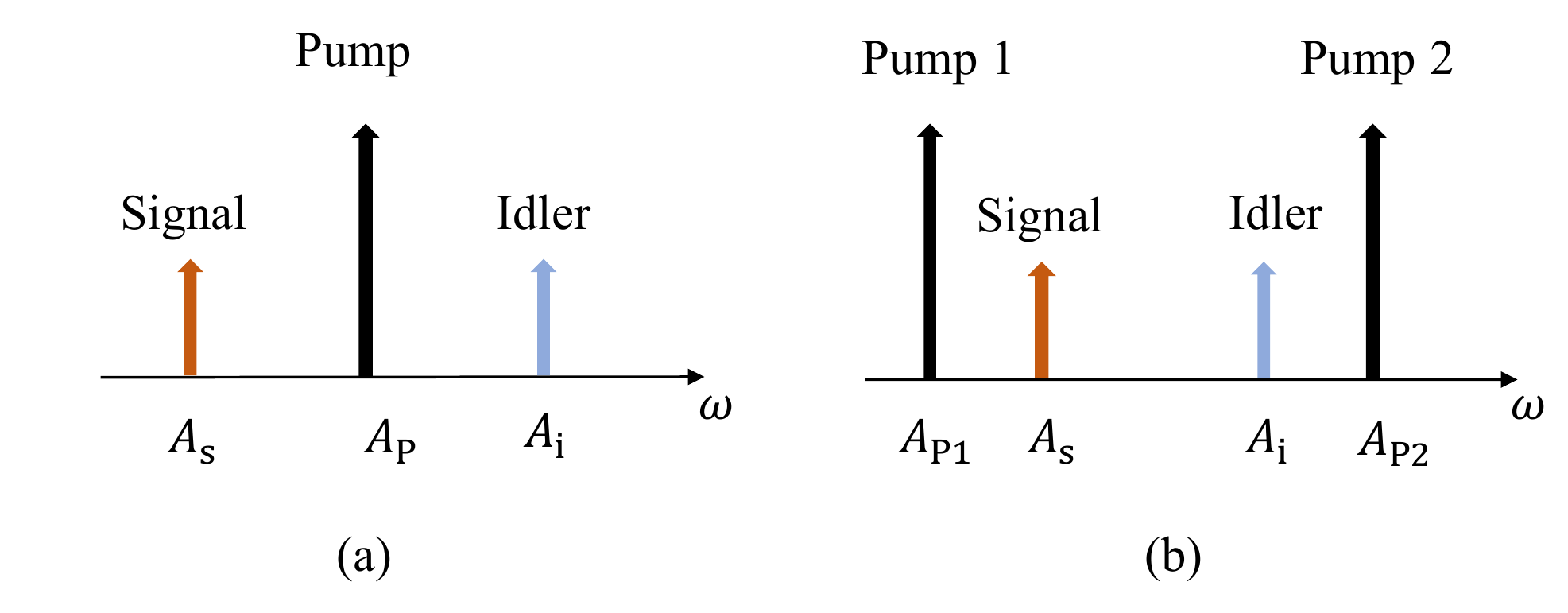}
\caption{\textcolor{black}{(a) Single-pump and (b) dual-pump FOPAs.}}\label{fig0}
\end{figure}

\textcolor{black}{\section{Basic aspects of PSAs and their applications}\label{sec.1N1}
Parametric gain originates from nonlinear wave-mixing between several waves due to the modulation of the medium parameters, mainly the refractive index. It can be obtained through three-wave mixing (TWM) in $\chi^{(2)}$ nonlinear materials, e.g., KTP-crystals \cite{levenson1993quantum}, periodically poled lithium niobate (PPLN), either in bulk or waveguide architecture \cite{lee2009phase, umeki2013line}, or through four-wave mixing (FWM) in $\chi^{(3)}$ non-linear materials such as highly nonlinear fibers (HNLFs) \cite{slavik2010all, tong2011towards} or silicon waveguides \cite{da2014phase}. Parametric gain through FWM mechanism is obtained by the amplification of signal and idler waves by two strong pump waves, in a manner such that the energy conservation is satisfied. An amplifier utilizing parametric gain in optical fibers is called a fiber optical parametric amplifier (FOPA). Depending on how the frequencies of the interacting pump, signal and idler waves are chosen at the input of the FOPA, two schemes are usually preferred, as shown in Fig. \ref{fig0}, namely  (a) the single-pump FOPA, and (b) the dual-pump FOPA. Both configurations have been widely investigated theoretically \cite{marhic1996broadband,floridia2003optimization, boggio2009155, boggio2002demonstration, kim2001gain}, and experimentally \cite{torounidis2006fiber, torounidis2007broadband, boggio2005double, boggio2006broadband, tang2005line}. Comparing the two configurations, it has been shown that with appropriate pump frequency allocation and optimized zero-dispersion frequency fluctuations, a dual-pump FOPA can achieve a flatter gain profile over a wider frequency range with less power for each pump, compared to the single-pump case \cite{yaman2004impact, liu2006optimization, radic2003two}. This feature makes dual-pump FOPAs very attractive for optical telecommunication applications \cite{torounidis2003amplification}.}

\textcolor{black}{In addition, FOPAs are capable of operating either as PSAs or PIAs. If the idler is injected at the input along with the signal and pump waves (see Fig.\;\ref{fig0}), the FOPA operates as a PSA, for which the amplifier gain depends on the relative phase between the interacting waves. However, in absence of the idler at the input, the FOPA operates as a PIA,  where the gain is independent of the initial phase of the signal. Most conventional optical amplifiers such as EDFAs, Raman amplifiers, semiconductor optical amplifiers (SOA) are PIAs.}
\textcolor{black}{Owing to the unique phase-sensitive property of PSAs, a 6 dB gain improvement compared with PIAs was demonstrated \cite{tong2011towards,tong2010phase, tong2011ultralow}.}
\textcolor{black}{Aside from the capability of amplifying the signal with large gain over the desired optical frequency range, the PSA is able to amplify one quadrature of the signal field wave, while de-amplifying the orthogonal quadrature \cite{levenson1993reduction}. This leads to the fact that the SNR on the amplified quadrature is not  degraded, providing thus an amplification of the signal with a quantum-limited NF of 0 dB, compared to 3 dB for conventional PIAs.
This squeezing property has an essential importance in several applications e.g., pulse reshaping \cite{imajuku1999pulse}, soliton-soliton interaction \cite{kath1998long}, and quantum noise suppression \cite{deutsch1994reduction}.
In practical implementations, the main challenge for the realization of PSAs is the need of phase locking among the interacting waves, especially if the waves are widely separated in frequency. Other challenges are polarization alignment, time synchronization of the interacting waves, and suppression of stimulated Brillouin scattering (SBS).
Nowdays, the basic features of PSAs are well understood, and many challenges have been solved from the application point of view, while others still remaining to be solved.
}

\section{Evolution of the field and basic principles}\label{sec.2}
In this section, we first recall the classical and quantum equations that describe the propagation of the field mode in fibres, and the definitions associated with the NF. We then discuss the linear input-output relations for different models.

\subsection{Propagation of classical and quantum fields }
fibre optical parametric amplifiers (FOPAs), based on Kerr non-linear interaction in optical fibres, use four-wave mixing (FWM) to achieve amplification. The gain mechanism is based on energy exchanges between several interacting optical fields in a fully elastic manner, i.e., without energy storage inside the medium. 

The total electric field propagating through the optical fibre along the $z$ direction is defined as:
\begin{equation}\label{eq1}
E(x,y,z,t) = \frac{f(x,y)}{2} \bigl\{A(z,t) e^{-i\left(\omega_{0} t- \beta(\omega_{0}) z\right)} + c.c.\bigr\}   \,,
\end{equation}
where $c.c.$ denotes the complex conjugate, $f(x,y)$ represents the transverse mode profile, $A(z,t)$ is the slowly varying complex amplitude of the propagating field and $\beta(\omega)$ is the propagation constant at frequency $\omega$. Its frequency variation is usually described by a Taylor expansion around the center frequency $\omega_{0}$ \cite{Agrawal}.


First, we have assumed in Eq.\,(\ref{eq1}) that the electrical field maintains its polarisation along the fibre length and can thus be considered as a scalar quantity. Second, we move to the reference frame that travels at the group velocity $v_{g}$ of the wave field, leading to a change of time variable $T=t-z/v_{g}$. Then, starting from Maxwell's equation, the propagation of the field envelope in the fibre is shown to be governed by the nonlinear Schr\"odinger equation (NLSE) \cite{Agrawal}:
\begin{equation}\label{eq2}
    \frac{\alpha}{2} A+ \frac{\partial A}{\partial z} + i \sum_{k=2}^{n} \frac{\beta_{k}}{k!}\frac{\partial^{k} A}{\partial T^{k}} = i\gamma |A|^{2} A \,,
\end{equation}
where $\beta_{k}$ is the $k$-th order derivative of $\beta$ with respect to $\omega$ taken at $\omega_0$ and $\gamma$ and $\alpha$ is the nonlinear coefficient and attenuation of the fibre.

The time dependence in Eq.\,(\ref{eq2}) can be neglected if the light spectrum is narrow enough to allow dispersion to be neglected, leading to:
\begin{equation}\label{eq3}
\frac{\alpha}{2} A + \frac{\partial A}{\partial z}= i \gamma |A|^{2} A\,.
\end{equation} 

In Heisenberg representation for quantized fields, the slowly varying amplitude $A$ of the field is replaced by the slowly varying annihilation operator $\hat{a}$, whose evolution satisfies the following equation \cite{Perinova}:
\begin{equation} \label{eq4}
\frac{d\hat{a}}{d t} = \frac{1}{i \hbar}[\hat{a},\hat{H}]\ ,
\end{equation}
where $\hat{H}$ is the Hamiltonian. In order to account for the spatial evolution of the field over $z$, Eq. (\ref{eq4}) is modified by replacing $dt$ by $\frac{n}{c} dz$, with $n$ is the effective refractive index \cite{huttner1990quantum}. 

It is worth noting that the Heisenberg equation (\ref{eq4}) is the quantum counterpart of the classical coupled-wave equation (\ref{eq3}). The transition from the classical theory to quantum mechanics can be readily achieved, by replacing the classical amplitude $A$ and its complex conjugate $A^{*}$ by the annihilation and creation operators $\hat{a}$ and     $\hat{a}^{\dagger}$, respectively \cite{V.Perinova, ben1992concept, Marhicbook}.

In the present paper, the classical equation (\ref{eq3}) will be used to derive a linear relation between the input and output modes of the field, which will be used afterwards to calculate the noise figure from the corresponding quantum operators. 
\subsection{Noise figure}
The noise added to the signal by the amplifier can be quantified using the noise figure NF \cite{Baney} defined as:
\begin{equation} \label{eq5}
\mathrm{NF}=\frac{\mathrm{SNR}_{\mathrm{in}}}{\mathrm{SNR}_{\mathrm{out}}}\ ,
\end{equation}
where $\mathrm{SNR}_{\mathrm{in}}$ and $\mathrm{SNR}_{\mathrm{out}}$ are the input and output signal-to-noise ratios, respectively. 

In terms of photon number, Eq. (\ref{eq5}) becomes:
\begin{equation}\label{eq6}
\mathrm{NF}= \frac{\langle N\rangle_{\mathrm{in}}^{2}\langle\Delta N^{2}\rangle_{\mathrm{out}}}{\langle N\rangle_{\mathrm{out}}^{2}\langle\Delta N^{2}\rangle_{\mathrm{in}}} \,,
\end{equation}
where  $\langle N\rangle_{\mathrm{in}}$ and $\langle N\rangle_{\mathrm{out}}$ are the input and output mean photon numbers, respectively, and $\langle\Delta N^{2}\rangle_{\mathrm{in}}$ and $\langle\Delta N^{2}\rangle_{\mathrm{out}}$ are the corresponding variances.

The average number of photons is expressed as: 
\begin{equation}\label{eq7}
\langle N\rangle = \langle \hat{a}^{\dagger}\hat{a} \rangle =\bra{\psi}\hat{a}^{\dagger}\hat{a}\ket{\psi} \,,
\end{equation}
where $\ket{\psi}$ is an initial state. Its fluctuation is given by the variance of $\langle N\rangle$ according to:
\begin{equation}\label{eq8}
\langle\Delta N^{2}\rangle = \langle N^{2}\rangle - \langle N\rangle^{2} \,.
\end{equation}

By assuming that the initial state is a coherent state, one can take $|\psi\rangle =|\alpha\rangle $, with $\alpha=|\alpha| e^{i\theta}$ is a complex number, $|\alpha|$ and $\theta$ are the amplitude and phase of the state $\ket{\alpha}$. 

In the following sections, the NF of Eq. (\ref{eq6}) will be calculated by evaluating the mean and variance of the photon number at the input and output of the amplifier.  Thus, a linear system linking the input and output modes is needed.

\subsection{Input-output linear relations}\label{sec.2.3}
As mentioned above, the nonlinear propagation of light modes in the fibre is well described by Eq.\,(\ref{eq3}), which can be solved analytically and/or numerically depending on the number of interacting modes. When  analytical solutions can be derived, input-output linear relations can be thus obtained, which can be used to calculate the noise figure using Eq.\,(\ref{eq6}). In the case of a dual-pump amplifier, Figure\,\ref{fig1} shows the three different configurations that we consider in this paper. Since the two pumps will be treated classically, we denote these configurations by the number of quantum modes that they exhibit.  They are thus called respectively the two-mode model [Fig.\,\ref{fig1}(a)], the four-mode model [Fig.\,\ref{fig1}(b)], and
the three-wave toy model [Fig.\,\ref{fig1}(c)]. We  discuss these schemes separately below, and find out under which conditions analytical solutions can be found for the input-output relations.
\subsubsection{Two-mode amplifier}\label{sec.2.3.1}
\begin{figure*}
\includegraphics[width=\textwidth]{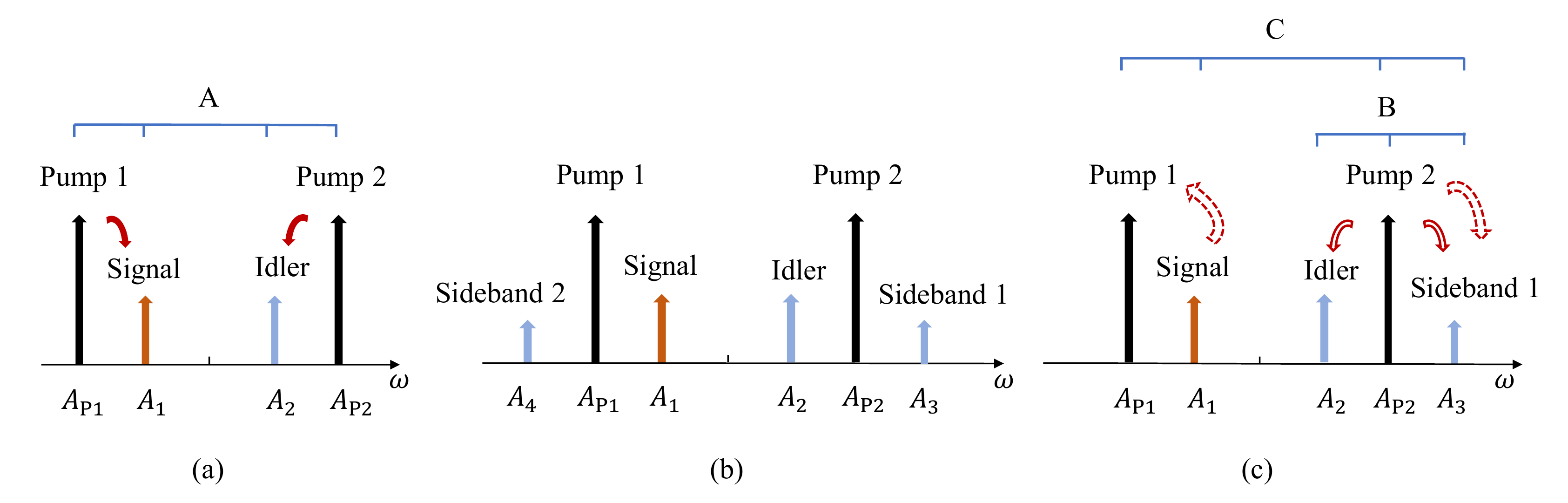}
\caption{\textcolor{black}{Schematic representation of different parametric amplifier models with non-degenerate pumps and non-degenerate signal and idler; (a) the 4-wave model, also called two(-quantum)-mode model, (b) the 6-wave model, also called four-mode model, (c) the three-mode model. Process A: annihilation of one pump 1 photon and one pump 2 photon, with creation of one signal photon and one idler photon.  Process B: Annihilation of two photons from pump 2 and creation of one idler photon and one sideband 1 photon. Process C: annihilation of one pump 2 photon and one signal photon and creation of one sideband 1 photon and one pump 1 photon. Modes $1$ and $2$ refer to the signal and idler, respectively, and modes $3$ and $4$ to sidebands $1$ and $2$, respectively. The pumps are not counted as quantum modes since they are treated as classical fields. When $A_{2} = 0$ at the input, the parametric amplifier operates as a PIA.}}\label{fig1}
\end{figure*} 
Figure \ref{fig1}(a) corresponds to the case where the non-degenerate signal and idler waves are amplified by FWM interaction with two strong pump waves. If pump depletion and fibre attenuation can be neglected, this situation becomes the so-called two-mode model, for which exact solutions have been extensively investigated in the literature \cite{Agrawal, Marhicbook}. When the pumps are treated as constant classical fields, it can be summarised as the following input-output linear relation:
\begin{equation}\label{eq9}
\begin{pmatrix}
      \hat{b}_{1}\\
      \hat{b}_{2}^{\dagger} \\
\end{pmatrix}
= 
\begin{pmatrix}
     \mu_{11} &  \mu_{12}  \\
     \mu_{21} &  \mu_{22} \\
\end{pmatrix}
\
\begin{pmatrix}
      \hat{a}_{1}\\
      \hat{a}_{2}^{\dagger} \\
\end{pmatrix}\ ,
\end{equation}
where $A=(\hat{a}_{1},\hat{a}_{2}^{\dagger})^{T}$ and $B=(\hat{b}_{1},\hat{b}_{2}^{\dagger})^{T}$ are vector operators for the input and output modes, respectively. The coefficients $\mu_{jk}$ of the  transfer matrix $M=[\mu_{jk}]$, with $j, k=1, 2$, are in general complex.

By evaluating the commutator $[\hat{b}_{j}, \hat{b}_{j}^{\dagger}] = 1$ for each mode $j$, the coefficients $\mu_{jk}$ have to satisfy the following relations:
\begin{align}\label{eq10}
\begin{cases}
 |\mu_{11}|^{2} - |\mu_{12}|^{2} = 1 \, , \\
 - |\mu_{21}|^{2} + |\mu_{22}|^{2} = 1 \, . \\
\end{cases}
\end{align}
These relations imply that the total number of photons is conserved. From a classical point view, Eqs.\,(\ref{eq10}) are known as the Manley-Rowe relations \cite{karlsson2015transmission}.

When pump depletion, fibre attenuation, and dispersion are neglected, the explicit expressions of the $\mu_{jk}$'s are given by \cite{Marhicbook}:
\begin{equation}\label{eq11}
 \left\{
    \begin{array}{ll}
       \mu_{11} = \cosh(\sqrt{3}\gamma P z) + \frac{i}{\sqrt{3}} \sinh(\sqrt{3}\gamma P z) \,,  \\
        \mu_{12} = \frac{2 i}{\sqrt{3}} \sinh(\sqrt{3}\gamma P z) \,,  \\
       \mu_{21} =-\frac{2 i}{\sqrt{3}}\sinh(\sqrt{3}\gamma P z) \,, \\
        \mu_{22} =  \cosh(\sqrt{3}\gamma P z) -\frac{i}{\sqrt{3}}\sinh(\sqrt{3}\gamma P z) ,
    \end{array}
\right.
\end{equation}
where $P\equiv P_{1}=P_{2}$ is the power of the two pumps. This two-mode model has been extensively investigated in order to predict and understand the gain and noise performances of the two-mode parametric amplifier \cite{Ferrini, Inoue,vedadi2006theoretical, mckinstrie2002parametric}. 
\subsubsection{Four-mode amplifier}\label{sec.2.3.2}
In practice, launching inside the fibre two strong pump waves with a signal and idler waves, as shown in Fig.\,\ref{fig1}(a), can lead to the creation of many extra waves by cascaded FWM interaction. The two first waves generated by FWM are the so-called sidebands 1 and 2 shown in Fig.\,\ref{fig1}(b), which are created symmetrically to the idler and signal with respect to  pump 2 and to pump 1, respectively. The six coupled-wave equations governing the evolution along the fibre of the amplitudes of these six interacting waves can be found in \cite{Marhicbook, Vedadi}. Their analytical solutions are complex and still unsolved in the general case, in contrast with the two-mode model. However, some attempts have been performed along this direction \cite{Colin, Marhicbook, Vedadi} and, under some assumptions, analytical solutions have been derived. If we suppose that the two pumps are not depleted, the signal, idler, sideband 1 and sideband 2 constitute the so-called four-mode model. 
In Ref. \cite{Colin}, \emph{McKinstrie \& al.} have proposed an analytical solution for this four-mode model for the particular case where dispersion is neglected, leading to the following input-output linear relation:  
\begin{equation}\label{eq12}
\begin{pmatrix}
      \hat{b}_{1}\\
      \hat{b}_{2}^{\dagger} \\
      \hat{b}_{3} \\
      \hat{b}_{4}^{\dagger} \\
\end{pmatrix}
= 
\begin{pmatrix}
     \mu_{11} &  \mu_{12} &  \mu_{13} &  \mu_{14}\\
     \mu_{21} &  \mu_{22} &  \mu_{23} &  \mu_{24}\\
     \mu_{31} &  \mu_{32} &  \mu_{33} &  \mu_{34}\\ 
     \mu_{41} &  \mu_{42} &  \mu_{43} &  \mu_{44}\\ 
\end{pmatrix}
\
\begin{pmatrix}
      \hat{a}_{1}\\
      \hat{a}_{2}^{\dagger} \\
      \hat{a}_{3} \\
      \hat{a}_{4}^{\dagger} \\
\end{pmatrix}\ ,
\end{equation}
where indices $1$ and $2$ denote the signal and idler, respectively, and indices $3$ and $4$ refer to the sidebands $1$ and $2$ (see Fig.\,\ref{fig1}(b)). The transfer matrix $M$ containing coefficients $\mu_{jk}$ with $(j,k=1,..4)$ is given by \cite{Colin, Marhicbook}:
\begin{equation}\label{eq13}
M(z)=
\begin{pmatrix}
1+i\gamma P z \hspace*{0.3cm} &  2i\gamma P z \hspace*{0.3cm} & 2i\gamma P z \hspace*{0.3cm} & i\gamma P z \\
  -2i\gamma P z  \hspace*{0.3cm} &  1-i\gamma P z \hspace*{0.3cm} &  -i\gamma P z \hspace*{0.3cm} & -2i\gamma P z\\
    2i\gamma P z \hspace*{0.3cm} & i\gamma P z \hspace*{0.3cm} & 1+i\gamma P z \hspace*{0.3cm} & 2i\gamma P z \\
    -i\gamma P z \hspace*{0.3cm} & -2i\gamma P z \hspace*{0.3cm} & -2i\gamma P z \hspace*{0.3cm} & 1-i\gamma P z
\end{pmatrix}\ .
\end{equation}

Similarly to Eq.\,(\ref{eq10}), the coefficients $\mu_{1k}$ that correspond to mode $1$ have to satisfy the following condition:
\begin{equation}\label{eq14}
 |\mu_{11}|^{2} - |\mu_{12}|^{2} + |\mu_{13}|^{2} - |\mu_{14}|^{2} = 1 \, , 
\end{equation}
with similar conditions for modes $j=2,..4$.

This four-mode model will be used in Section\,\ref{sec.4} below in order to predict the gains and noise figures of the four interacting modes.
\subsubsection{A toy model: the three-mode amplifier}
The four-mode model of the preceding subsection leads to much more complicated calculations than the simple two-mode model of subsection \ref{sec.2.3.1}. An intermediate toy model can be designed if we neglect one of the sidebands, for example sideband 2, in the scheme of Fig.\,\ref{fig1}(b), leading to the scheme of  Fig.\,\ref{fig1}(c). Of course, this model does not correspond to any realistic physical situation, but we will use it as an intermediate step between the two-mode model and more general and useful situations in the quantum noise calculations. 

Reducing Eq.\,(\ref{eq12}) to this situation, we describe the three-mode model by the following input-output linear relation:
\begin{equation}\label{eq15}
\begin{pmatrix}
      \hat{b}_{1}\\
      \hat{b}_{2}^{\dagger} \\
      \hat{b}_{3} \\
\end{pmatrix}
= 
\begin{pmatrix}
     \mu_{11} &  \mu_{12} &  \mu_{13} \\
     \mu_{21} &  \mu_{22} &  \mu_{23} \\
     \mu_{31} &  \mu_{32} &  \mu_{33} \\ 
\end{pmatrix}
\
\begin{pmatrix}
      \hat{a}_{1}\\
      \hat{a}_{2}^{\dagger} \\
      \hat{a}_{3} \\
\end{pmatrix}
\end{equation}
where 
$j,k=1,2,3$ refer to the signal, idler, and sideband $1$, respectively (see Fig.\,\ref{fig1}(c)). Notice that Eq.\,(\ref{eq14}) reduces, with $\mu_{14}=0$, to:
\begin{equation}\label{eq16}
|\mu_{11}|^{2} - |\mu_{12}|^{2} + |\mu_{13}|^{2} = 1 \,.
\end{equation}

The explicit expressions of the $\mu_{jk}$'s are not given here because, as stated above, this three-mode model does not correspond to any realistic physical situation in a dual-pump parametric amplifier. Nevertheless, we will use it in the following section as an intermediate step in the calculation of the noise figure.

\section{Quantum treatment of the noise figure}\label{sec.3}
In this section, we adopt the quantum approach introduced by earlier researchers \cite{Caves,Levenson,louisell1961quantum,haus1962quantum, gordon1963quantum} to evaluate the noise figure defined by Eq.\,(\ref{eq6}). This approach treats the electric field of each mode as an operator, leading to the appearance of an intrinsic quantum noise accompanying the gain mechanism.
This leads to an extra photon noise, which is evaluated by means of Eq.\,(\ref{eq8}), while the change in average photon number in the amplification process is obtained using Eq.\,(\ref{eq7}). By the use of this quantum approach, our motivation is to derive a generalised expression of the noise figure for many-mode models. However, before we move on to the general case, we discuss the noise properties of the particular cases presented in Section \ref{sec.2.3}. Indeed, in the following, we first recall the familiar NF calculation for the two-mode model. Second, as pointed out earlier, we  make use of the three-mode model, as an intermediate step, to analyse the noise figure when three interacting modes are accounted for. Then, by comparing the results of both models, we deduce the general NF expression that works for many-mode models. The noise figure for the case of the four-mode model is discussed later in Section \ref{sec.4}.  

We underline the fact that the derivation of the noise figure is performed below for both phase insensitive and phase sensitive amplification. For the case of the PIA, the results of our derivation can be compared with the one already discussed in Ref.\,\cite{Colin}. On the contrary, the derivation in the case of the PSA has no equivalent in the literature.

\subsection{Two-mode amplifier}\label{sec.3.1}
We make use of the input-output relation (\ref{eq9}) to evaluate the noise figure of the two-mode model (Fig.\,\ref{fig1}(a)) for both PIA and PSA configurations. In terms of quantum noise, this model has been thoroughly investigated in the literature \cite{C.J.McKinstrie, Ferrini,Inoue}. Here we remind the NF calculation in this case for the sake of comparison with the other models discussed in this paper.
\subsubsection{Phase insensitive amplification}
When only the two pumps and the signal are present at the input of the fibre, the two-mode model in Fig.\,\ref{fig1}(a) operates as a PIA. Thus, the initial state is the tensor product of a coherent state of complex amplitude $\alpha_{1}=|\alpha_{1}| e^{i\theta_{1}}$ for the signal and the vacuum state for the idler, i.e.,  $|\psi>=|\alpha_{1},0> $.

Since we assume that the initial signal state is a coherent state, the photon number has a Poissonian distribution, for which the  variance is equal to the mean photon number:
\begin{equation}\label{eq17}
\langle N_{1}\rangle_{\mathrm{in}} = \langle\Delta N_{1}^{2}\rangle_{\mathrm{in}} = |\alpha_{1}|^{2} \,.
\end{equation}
The output mean photon number of mode $j$ is defined as: 
\begin{equation}\label{eq18}
    \langle N_{j}\rangle _{\mathrm{out}}= \langle \hat{b}_{j}^{\dagger}\hat{b}_{j} \rangle _{\mathrm{out}} \, ,
\end{equation}
Using the input-output relation (\ref{eq9}), Eq.\,(\ref{eq18}) leads to the following expressions for the output mean photon numbers for the signal and idler:
\begin{align}
\langle N_{1}\rangle_{\mathrm{out}}&= |\mu_{11}|^{2}  |\alpha_{1}|^{2} + |\mu_{12}|^{2} \, , \label{eq19} \\
\langle N_{2}\rangle_{\mathrm{out}}&= |\mu_{21}|^{2} \left(1 + |\alpha_{1}|^{2}\right) \, \label{eq20}.
\end{align}
\textcolor{black}{Equation (\ref{eq19}) shows that the mean photon number of the signal consists of two components: the first term corresponds to the amplified signal photons, and the second term corresponds to vacuum fluctuations, stemming from the interaction between the signal and idler by means of the two-mode process A shown in Fig.\,\ref{fig1}(a). The vacuum fluctuation term $1\times|\mu_{12}|^{2}$ in Eq.\,(\ref{eq19}) is due to the so-called parametric fluorescence phenomenon,  existing even when both the signal and idler are absent at the input as shown by Eqs.\ \,(\ref{eq19}) and (\ref{eq20}). In Eq.\,(\ref{eq20}), the output mean photon number for the idler does not contain any amplified idler term due to the absence of  idler power at the input. However, it depends on the input signal photon number $|\alpha_{1}|^{2}$ through  process A (see Fig.\,\ref{fig1}(a)), and includes a parametric fluorescence term in the form $1\times|\mu_{21}|^{2}$.}

The variances of the output signal and idler are defined by Eq.\,(\ref{eq8}) and expressed as:
\begin{align}
\langle\Delta N_{1}^{2}\rangle_{\mathrm{out}}&= |\mu_{11}|^{4} |\alpha_{1}|^{2} + |\mu_{11}\mu_{12}|^{2} \left( 1 + |\alpha_{1}|^{2} \right) \, ,\label{eq21}\\
\langle\Delta N_{2}^{2}\rangle_{\mathrm{out}}& = |\mu_{21}|^{4} |\alpha_{1}|^{2} + |\mu_{21}\mu_{22}|^{2} \left( 1 + |\alpha_{1}|^{2} \right) \, .\label{eq22}
\end{align}

\textcolor{black}{As an example, we detail the physical interpretation of Eq. (\ref{eq21}), as follows: the first term comes from the amplification of the input signal photons. The second term $1\times|\mu_{11}\mu_{12}|^{2} $ is due to the  parametric fluorescence in the two modes. Finally, the third term $|\mu_{11}\mu_{12}|^{2} |\alpha_{1}|^{2} $ results from the nonlinear interaction between the input signal and the quantum fluctuations of  the idler. }

\textcolor{black}{In the following, all the terms due to parametric fluorescence only will be neglected in the calculation of the noise figure, by assuming that $|\alpha_1|^{2},|\alpha_2|^{2} \gg 1$.}

The NF for the two interacting modes is calculated using Eq.\,(\ref{eq6}). If the mode $j$ is present at the fibre input, like the signal ($j=1$) in the present case, we consider the same mode at the input and output of the amplifier, leading to:
\begin{equation} \label{eq23}
\mathrm{NF}_{1} =  \frac{\langle  N_{1}\rangle_{\mathrm{in}}^{2}}{\langle\Delta N_{1}^{2}\rangle_{\mathrm{in}}} \cdot \frac{\langle\Delta N_{1}^{2}\rangle_{\mathrm{out}}}{\langle N_{1}\rangle_{\mathrm{out}}^{2}}\ .
\end{equation}
However, in the case where we want to evaluate the amount of noise injected in a mode $j$ that was not fed at the fibre input, like the idler ($j=2$) in the present case, we can no longer use Eq. (\ref{eq6}) because the input signal-to-noise ratio is zero. Following Refs.~\cite{Jopson, mckinstrie2003parametric}, we thus generalise the definition of Eq. (\ref{eq6}) by normalising the output signal-to-noise ratio of the considered mode $j$ to the input signal-to-noise ratio for the signal:
\begin{equation} \label{eq23N1}
\mathrm{NF}_{j} =  \frac{\langle  N_{1}\rangle_{\mathrm{in}}^{2}}{\langle\Delta N_{1}^{2}\rangle_{\mathrm{in}}} \cdot \frac{\langle\Delta N_{j}^{2}\rangle_{\mathrm{out}}}{\langle N_{j}\rangle_{\mathrm{out}}^{2}}\ .
\end{equation}
%

As an example, by substituting Eqs. (\ref{eq17}, \ref{eq19}, \ref{eq21}) into Eq.\,(\ref{eq23}), and by supposing a relatively strong input signal intensity $|\alpha_{1}|^{2}\gg 1$, the signal NF in the case of the two-mode PIA is found to be equal to:
\begin{equation}\label{eq24}
    \mathrm{NF}_{\mathrm{1,PIA}} = 1 +  \frac{|\mu_{12}|^{2}}{|\mu_{11}|^{2}}\, .
\end{equation}
Besides, using Eqs.\,(\ref{eq9}, \ref{eq10}), the phase insensitive signal gain in this case is:
\begin{equation}\label{eq25}
    \textcolor{black}{G_{\mathrm{1,PIA}} =|\mu_{11}|^{2}=|\mu_{12}|^{2}+1 }\, .
\end{equation}
\textcolor{black}{Hence, Eq.\,(\ref{eq24}) can be rewritten in the following form: 
\begin{equation}\label{eq26}
    \mathrm{NF}_{\mathrm{1,PIA}}=2-\frac{1}{G_{\mathrm{1,PIA}}}\, ,
\end{equation}
which is the well-known result for the two-mode PIA noise figure, approaching 3 dB  in the high-gain regime.}

\subsubsection{Phase sensitive amplification}
Let us now consider the case where the amplifier of  Fig. \ref{fig1}(a) operates as a PSA, i.e. when both the signal and idler are present at the input. The input field  state is thus $|\psi\rangle=|\alpha_{1},\alpha_{2}\rangle $, where $\alpha_{1}=|\alpha_{1}| e^{i\theta_{1}}$ and $\alpha_{2} = |\alpha_{2}| e^{i\theta_{2}}$.


Similarly to the PIA case, Eq.\,(\ref{eq18}) leads to the following output average photon numbers for the signal and idler:
\begin{multline}\label{eq27}
\langle N_{1}\rangle_{out} = |\mu_{11}|^{2} |\alpha_{1}|^{2} + |\mu_{12}|^{2} \bigl( 1 + |\alpha_{2}|^{2}\bigr) \\
+ \bigl(\mu_{11}\mu_{12}^{*}\alpha_{1}\alpha_{2} +\mu_{11}^{*}\mu_{12}\alpha_{1}^{*}\alpha_{2}^{*} \bigr)\, ,
\end{multline}
\begin{multline}\label{eq28}
\langle N_{2}\rangle_{out} = |\mu_{21}|^{2} ( 1 + |\alpha_{1}|^{2} ) + |\mu_{22}|^{2} |\alpha_{2}|^{2} \\
+ \bigl(\mu_{21}\mu_{22}^{*}\alpha_{1}\alpha_{2} +\mu_{21}^{*}\mu_{22}\alpha_{1}^{*}\alpha_{2}^{*} \bigr) \, .
\end{multline}
Their variances are found to be:
\begin{multline}\label{eq29}
\langle\Delta N_{1}^{2}\rangle_{out} = |\mu_{11}|^{4} |\alpha_{1}|^{2} + |\mu_{12}|^{4} |\alpha_{2}|^{2} \\
+ |\mu_{11}\mu_{12}|^{2} \bigl(1+|\alpha_{1}|^{2} + |\alpha_{2}|^{2} \bigr) +\bigl(|\mu_{11}|^{2}+|\mu_{12}|^{2}\bigr) \\
\bigl(\mu_{11}\mu_{12}^{*}\alpha_{1}\alpha_{2} +  \mu_{11}^{*}\mu_{12}\alpha_{1}^{*}\alpha_{2}^{*} \bigr) \, ,
\end{multline}
\begin{multline}\label{eq30}
\langle\Delta N_{2}^{2}\rangle_{out} = |\mu_{21}|^{4} |\alpha_{1}|^{2} + |\mu_{22}|^{4} |\alpha_{2}|^{2} \\
+ |\mu_{21}\mu_{22}|^{2} \bigl(1+|\alpha_{1}|^{2} + |\alpha_{2}|^{2}\bigr) + \bigl(|\mu_{21}|^{2}+|\mu_{22}|^{2}\bigr) \\
\bigl(\mu_{21}\mu_{22}^{*}\alpha_{1}\alpha_{2} + \mu_{21}^{*}\mu_{22}\alpha_{1}^{*}\alpha_{2}^{*} \bigr) \, .
\end{multline}

\textcolor{black}{Equations (\ref{eq27}-\ref{eq30}) can be interpreted in a similar way as equations (\ref{eq19}-\ref{eq22}), the difference relying in the presence of the idler at the input, which gives rise to new terms like $|\mu_{22}|^{2} |\alpha_{2}|^{2}$ in Eq.(\ref{eq28}) and $|\mu_{11}\mu_{12}|^{2} |\alpha_{2}|^{2}$ in Eq.(\ref{eq29}). These terms merely originate from the amplification of input idler photons, and the interaction between the idler photons and the   fluctuations of the input signal, respectively. Another important difference is the presence of the term ($\mu_{j1}\mu_{j2}^{*}\alpha_{1}\alpha_{2} +\mathrm{c.c.})$, which depends on the input signal phase, confirming thus the phase dependence property of PSAs, as compared to PIAs.}

The NF can be deduced from Eq.\,(\ref{eq6}) for the signal and idler. For brevity, we reproduce here only the NF expression for the signal assuming that $|\alpha_{1}|^{2} = |\alpha_{2}|^{2}$, with $|\alpha_{1,2}|^{2} \gg 1$. Taking $\mu_{11}=|\mu_{11}| e^{i\theta_{11}}$, $\mu_{12}=|\mu_{12}| e^{i\theta_{12}}$ and substituting Eqs.\,(\ref{eq17}, \ref{eq27}, \ref{eq29}) into Eq.\,(\ref{eq23}), the signal NF is found to be given by:
\begin{multline}\label{eq31}
\mathrm{NF}_{\mathrm{1,PSA}} = \Bigl\{ |\mu_{11}|^{4}+|\mu_{12}|^{4}+ 2 |\mu_{11}\mu_{12}|^{2}\\
+ 2 |\mu_{11}\mu_{12}| \bigl(\sum_{k=1}^{2}|\mu_{1k}|^{2}\bigr) \cos{\Theta} \Bigr\} 
\\
\times\frac{1}{\bigl(|\mu_{11}|^{2} + |\mu_{12}|^{2} + 2 |\mu_{11}\mu_{12}| \cos{\Theta}\bigr)^{2}} \, ,
\end{multline}
with $\Theta = \theta_{11}-\theta_{12}+\theta_{1}+\theta_{2}$ the relative phase between the pumps, signal and idler. The phases $\theta_{11}$ and $\theta_{12}$ indeed depend on the input phases of the pumps.

From the linear relation (\ref{eq9}), the phase sensitive gain for the signal is found to be:
\begin{equation}\label{eq32}
    G_{\mathrm{1,PSA}}= |\mu_{11}|^{2} + |\mu_{12}|^{2} + 2 |\mu_{11}\mu_{12}| \cos{\Theta} \, .
\end{equation}
Hence, using Eqs.\,(\ref{eq25}, \ref{eq32}), the signal NF of Eq.\,(\ref{eq31}) can be rewritten as:
\begin{equation}\label{eq33}
    \mathrm{NF}_{\mathrm{1,PSA}}= \frac{2G_{\mathrm{1,PIA}}-1}{G_{\mathrm{1,PSA}}}  \, ,
\end{equation}
which is the classical expression for a two-mode PSA.
\subsection{Three-mode amplifier}\label{sec.3.2}
Using the input-output relation (\ref{eq15}), we now evaluate the NF in the case when three interacting modes are considered (Fig.\,\ref{fig1}(c)). We follow the analysis of Section \ref{sec.3.1} and we consider as well the two amplifying scenarios.
\subsubsection{Phase insensitive amplification}
When the three-mode scheme of Fig.\,\ref{fig1}(c) operates as a PIA, the initial state is given as $|\psi\rangle=|\alpha_{1},0,0\rangle$. The idler and sideband $1$ are injected with vacuum.


The mean photon numbers at the amplifier output are evaluated using Eq.\,(\ref{eq18}), leading to:
\begin{align}
\langle N_{1}\rangle_{\mathrm{out}}&= |\mu_{11}|^{2}  |\alpha_{1}|^{2} + |\mu_{12}|^{2} \, , \label{eq34} \\
\langle N_{2}\rangle_{\mathrm{out}}&= |\mu_{21}|^{2} \bigl(1 + |\alpha_{1}|^{2}\bigr) + |\mu_{23}|^{2} \, \label{eq35} , \\
\langle N_{3}\rangle_{\mathrm{out}}&= |\mu_{31}|^{2} |\alpha_{1}|^{2} + |\mu_{32}|^{2} \label{eq36}.
\end{align}

\textcolor{black}{Compared to the two-mode model, in which only one fundamental parametric process  is considered (see Fig.\,\ref{fig1}(a)), the present toy model involves two extra processes (see Fig.\,\ref{fig1}(c)). Consequently, some new terms appear in equations (\ref{eq34}-\ref{eq36}) giving the mean output  photon numbers. }

\textcolor{black}{Let us interpret  these equations separately. First, the signal in the toy model  of Fig.\ref{fig1}(c) is amplified through the two processes labeled A and C. Process A is responsible for the term $ 1\times|\mu_{12}|^{2}$ in Eq.(\ref{eq34}), due to parametric fluorescence. Process C, also called frequency conversion, is a noiseless process, explaining why the term $1\times|\mu_{13}|^{2}$ is absent in Eq.\,(\ref{eq34}).}

\textcolor{black}{Second, the idler is amplified by means of two processes A and B.  Process A is responsible of the first term of Eq.(\ref{eq35}), where the value `1' refers to parametric fluorescence. Process B introduces the term $1\times|\mu_{23}|^{2}$, which comes from the vacuum fluctuations injected in  sideband $1$.}

\textcolor{black}{Third, sideband $1$ is amplified by the processes B and C. Process B gives rise to the term $1\times|\mu_{32}|^{2}$, which originates from the fluctuations of the idler. Conversely, the frequency conversion process C, which gives rise to the term $|\mu_{31}|^{2} |\alpha_{1}|^{2}$, does not contain any  fluctuation term, since it is a noiseless process.}

The output variances are expressed as:
\begin{align}
\langle\Delta N_{1}^{2}\rangle_{\mathrm{out}} =& |\mu_{11}|^{4} |\alpha_{1}|^{2} + |\mu_{11}\mu_{12}|^{2} \bigl( 1 + |\alpha_{1}|^{2} \bigr)\nonumber\\
&+ |\mu_{11}\mu_{13}|^{2}|\alpha_{1}|^{2} + |\mu_{12}\mu_{13}|^{2} \, ,\label{eq37}\\
\langle\Delta N_{2}^{2}\rangle_{\mathrm{out}} =& |\mu_{21}|^{4} |\alpha_{1}|^{2} + |\mu_{21}\mu_{22}|^{2} \bigl( 1 + |\alpha_{1}|^{2} \bigr)\nonumber\\
&+|\mu_{21}\mu_{23}|^{2}|\alpha_{1}|^{2} + |\mu_{22}\mu_{23}|^{2} \, ,\label{eq38}\\
\langle\Delta N_{3}^{2}\rangle_{\mathrm{out}} =& |\mu_{31}|^{4} |\alpha_{1}|^{2} + |\mu_{31}\mu_{32}|^{2} \bigl( 1 + |\alpha_{1}|^{2} \bigr)\nonumber\\
&+ |\mu_{31}\mu_{33}|^{2}|\alpha_{1}|^{2} + |\mu_{32}\mu_{33}|^{2} \, .\label{eq39}
\end{align}

\textcolor{black}{These equations can be interpreted in a similar way as equations (\ref{eq21},\ref{eq22}). The difference here comes from the existence of new terms associated with sideband $1$ and its interaction with the other modes.}

We notice that Equations (\ref{eq34}-\ref{eq39}) become more and more difficult to obtain analytically when the number of interacting modes increases. We thus developed the \emph{Mathematica} code given in the supplementary information to calculate the mean and the variance of the different photon numbers. This code can be used for an arbitrary number of interacting modes.

The signal NF for the three-mode PIA is obtained from Eqs.\,(\ref{eq23},\ref{eq34},\ref{eq37}) and reads, for $|\alpha_1|\gg1$, as:
\begin{equation}\label{eq40}
    \mathrm{NF}_{\mathrm{1,PIA}} = 1 +  \frac{|\mu_{12}|^{2}+|\mu_{13}|^{2}}{|\mu_{11}|^{2}} \, .
\end{equation}
\subsubsection{Phase sensitive amplification} \label{sec.3.2.2}
In this case, the field initial state is $|\psi>=|\alpha_{1},\alpha_{2},0> $. Following the same analysis as previously, the output mean photon numbers of the three modes are found to be:
\begin{multline}\label{eq41}
<N_{1}>_{\mathrm{out}} = |\mu_{11}|^{2} |\alpha_{1}|^{2} + |\mu_{12}|^{2} \bigl( 1 + |\alpha_{2}|^{2}\bigr) \\
+ \bigl(\mu_{11}\mu_{12}^{*}\alpha_{1}\alpha_{2} +\mu_{11}^{*}\mu_{12}\alpha_{1}^{*}\alpha_{2}^{*} \bigr)\, ,
\end{multline}
\begin{multline}\label{eq42}
<N_{2}>_{\mathrm{out}} = |\mu_{21}|^{2} ( 1 + |\alpha_{1}|^{2} ) + |\mu_{22}|^{2} |\alpha_{2}|^{2} \\
+|\mu_{23}|^{2} + \bigl(\mu_{21}\mu_{22}^{*}\alpha_{1}\alpha_{2} +\mu_{21}^{*}\mu_{22}\alpha_{1}^{*}\alpha_{2}^{*} \bigr) \, ,
\end{multline}
\begin{multline}\label{eq43}
<N_{3}>_{\mathrm{out}} = |\mu_{31}|^{2} |\alpha_{1}|^{2} + |\mu_{32}|^{2} \bigl( 1 + |\alpha_{2}|^{2} \bigr) \\
+\bigl(\mu_{31}\mu_{32}^{*}\alpha_{1}\alpha_{2} +\mu_{31}^{*}\mu_{32}\alpha_{1}^{*}\alpha_{2}^{*} \bigr) \, .
\end{multline}
And their output variances are expressed as :
\begin{multline}\label{eq44}
<\Delta N_{1}^{2}>_{\mathrm{out}} = |\mu_{11}|^{4} |\alpha_{1}|^{2} + |\mu_{12}|^{4} |\alpha_{2}|^{2} \\
+ |\mu_{11}\mu_{12}|^{2} \bigl(1+|\alpha_{1}|^{2} + |\alpha_{2}|^{2} \bigr) + |\mu_{11}\mu_{13}|^{2}|\alpha_{1}|^{2}  \\
+|\mu_{12}\mu_{13}|^{2} \bigl(1+ |\alpha_{2}|^{2} \bigr)
+ \bigl(|\mu_{11}|^{2}+|\mu_{12}|^{2}+|\mu_{13}|^{2}\bigr) \\
\bigl(\mu_{11}\mu_{12}^{*}\alpha_{1}\alpha_{2} + \mu_{11}^{*}\mu_{12}\alpha_{1}^{*}\alpha_{2}^{*} \bigr) \, ,
\end{multline}
\begin{multline}\label{eq45}
<\Delta N_{2}^{2}>_{\mathrm{out}} = |\mu_{21}|^{4} |\alpha_{1}|^{2} + |\mu_{22}|^{4} |\alpha_{2}|^{2} \\
+ |\mu_{21}\mu_{22}|^{2} \bigl(1+|\alpha_{1}|^{2} + |\alpha_{2}|^{2}\bigr) +  |\mu_{21}\mu_{23}|^{2}|\alpha_{1}|^{2}  \\ +|\mu_{22}\mu_{23}|^{2} \bigl(1 + |\alpha_{2}|^{2}\bigr) 
+ \bigl(|\mu_{21}|^{2}+|\mu_{22}|^{2}+|\mu_{23}|^{2}\bigr) \\
\bigl(\mu_{21}\mu_{22}^{*}\alpha_{1}\alpha_{2} + \mu_{21}^{*}\mu_{22}\alpha_{1}^{*}\alpha_{2}^{*} \bigr)  \, ,
\end{multline}
\begin{multline}\label{eq46}
<\Delta N_{3}^{2}>_{\mathrm{out}} = |\mu_{31}|^{4} |\alpha_{1}|^{2} + |\mu_{32}|^{4} |\alpha_{2}|^{2} \\
+ |\mu_{31}\mu_{32}|^{2} \bigl(1+|\alpha_{1}|^{2} + |\alpha_{2}|^{2} \bigr) + |\mu_{31}\mu_{33}|^{2}|\alpha_{1}|^{2} \\
+ |\mu_{32}\mu_{33}|^{2} \bigl(1 + |\alpha_{2}|^{2}\bigr)
+ \bigl(|\mu_{31}|^{2}+|\mu_{32}|^{2}+|\mu_{33}|^{2}\bigr) \\
\bigl(\mu_{31}\mu_{32}^{*}\alpha_{1}\alpha_{2}+ \mu_{31}^{*}\mu_{32}\alpha_{1}^{*}\alpha_{2}^{*} \bigr)\, .
\end{multline}

For brevity, we give here only the expression of the NF for the signal mode. Following the same assumption used for Eq.\,(\ref{eq31}), and substituting Eqs.\,(\ref{eq41},\ref{eq44}) into Eq.\,(\ref{eq23}), the signal NF is found to have the following expression:
\begin{multline}\label{eq47}
NF_{\mathrm{1,PSA}} = \Bigl\{ |\mu_{11}|^{4}+|\mu_{12}|^{4}+ 2 |\mu_{11}\mu_{12}|^{2} +  |\mu_{11}\mu_{13}|^{2} \\
+|\mu_{12}\mu_{13}|^{2}  
+ 2 |\mu_{11}\mu_{12}| \bigl(\sum_{k=1}^{3}|\mu_{1k}|^{2}\bigr) \cos{\Theta} \Bigr\} 
\\
\times\frac{1}{\bigl(|\mu_{11}|^{2} + |\mu_{12}|^{2} + 2 |\mu_{11}\mu_{12}| \cos{\Theta}\bigr)^{2}} \, ,
\end{multline}
where $\Theta = \theta_{11}-\theta_{12}+\theta_{1}+\theta_{2}$ is the input relative phase between the pumps, signal, and idler.

\subsection{Comparison between the models}
By comparing the results obtained for both of the two-mode (Figure \ref{fig1}(a) and Section \ref{sec.3.1}) and three-mode (Figure \ref{fig1}(c) and Section \ref{sec.3.2}) models, one can see that the increase of the number of involved modes gives rise to additional couplings between these modes, and thus leads to additional terms in the expressions of the  mean and variance of the output photon number.

From these expressions, one can find that for both PIA and PSA configurations, there are terms that depend on the $|\alpha_{j}|^{2}$'s, which are related to the contributions of the modes present at the fibre input. Additionally, there are some other terms, of the forms $1\times|\mu_{jk}|^{2}$ with $(j\neq k)$ and $1\times |\mu_{jl}\mu_{jm}|^{2}$, which are purely quantum contributions. These latter terms are associated with the generation of fields out of vacuum fluctuations in the initially empty modes. They depend of course on the type of involved operators, i.e. on whether these operators are of the same type (both creation or both annihilation operators), or of opposite types (one creation operator and one annihilation operator). As an example, in Eqs. (\ref{eq41}-\ref{eq46}) of Section \ref{sec.3.2.2}, the average and the variance of the output photon numbers of modes $1$ and $3$ are composed of quantum contributions that are related to mode $2$, whereas for mode $2$, these quantum contributions are related to modes $1$ and $3$. The difference between the PIA and PSA amplifying configurations lies in the presence of  terms of the form $\mu_{j1}\mu_{j2}^{*}\alpha_{1}\alpha_{2} +\mathrm{c.c.}$
in the expressions obtained for the PSA. These terms depend on the input phases of the pumps, signal, and idler. As a result, the NF expressions (\ref{eq31}) and (\ref{eq47}) for the PSA are found to depend on the relative phase $\Theta$.

We point out that, although the derivation of the NF for both two-mode and three-mode models remains achievable analytically, the situation gets much more complicated in the case of the four-mode model and, a fortiori, for more than four modes. The \emph{Mathematica} code given in the supplementary information can calculate the average and the variance of the output photon numbers for an arbitrary number of interacting modes. However, it does not give the final expression of the noise figure. For these reasons, we propose in the following a generalised expression of the noise figure for many-mode models, for both amplifying scenarios.
\subsection{Generalisation: Many-mode amplifier}\label{section4}
Based on the results of the previous subsections, we derive in this section a general expression of the NF for models involving many interacting modes. For this derivation, we assume that the many-mode amplifier is described by the following input-output linear relation:
\begin{equation}\label{eq48}
\begin{pmatrix}
      \hat{b}_{1}\\
      \hat{b}_{2}^{\dagger} \\
      \vdots \\
      \hat{b}_{n} \hspace{0.1cm}\text{or}\hspace{0.1cm} \hat{b}_{n}^{\dagger}
\end{pmatrix}
= \begin{pmatrix} 
\mu_{11} &  \mu_{12}  & \ldots & \mu_{1n}\\
\mu_{21}  &  \mu_{22} & \ldots & \mu_{2n}\\
\vdots & \vdots & \ddots & \vdots\\
\mu_{n 1}  &   \mu_{n 2}       &\ldots & \mu_{n n}
\end{pmatrix}
\
\begin{pmatrix}
      \hat{a}_{1}\\
      \hat{a}_{2}^{\dagger} \\
      \vdots \\
      \hat{a}_{n} \hspace{0.1cm}\text{or}\hspace{0.1cm} \hat{a}_{n}^{\dagger}
\end{pmatrix}\ ,
\end{equation}
where $A=(\hat{a}_{1},\hat{a}_{2}^{\dagger},..\hat{a}_{n} \hspace{0.1cm}\text{or}\hspace{0.1cm} \hat{a}_{n}^{\dagger})^{T}$ and $B=(\hat{b}_{1},\hat{b}_{2}^{\dagger},..\hat{b}_{n} \hspace{0.1cm}\text{or}\hspace{0.1cm} \hat{b}_{n}^{\dagger})^{T}$ are vector operators for the input and output mode operators, respectively. By convention, we choose mode number 1 as the signal. The last line of the input and output vectors is an annihilation or creation operator depending on the parity of the number of modes. The coefficients $\mu_{jk}$ of the  transfer matrix $M=[\mu_{jk}]$, with $j,k = 1, 2,..n$ are in general complex.

From Eqs.\,(\ref{eq10},\ref{eq14},\ref{eq16}), the coefficients $\mu_{jk}$ for $n$ interacting modes have to satisfy the following condition:
\begin{equation}\label{eq49}
\sum_{k=1}^{n} |\mu_{jk}|^{2} s_{jk} = 1 \, ,
\end{equation}
where $s_{jk}$ is equal to $1$ when the operators labeled by the integers $j$ and $k$ in Eq.\,(\ref{eq48}) are of the same type (both creation or both annihilation operators), or  to $-1$ when they are of opposite types (one creation operator and one annihilation operator). 

We take the input state of the many-mode field launched in the amplifier as: 
\begin{equation}\label{eq50}
|\psi\rangle = \sum_{j=1}^{n}|\alpha_{j},\alpha_{j+1},..., \alpha_{n} \rangle \, ,
\end{equation}
where $|\alpha_{j}\rangle=|0_{j}\rangle$, if no field is injected into mode $j$ at the amplifier input.

\subsubsection{Phase insensitive amplification}
From Subsections \ref{sec.3.1} and \ref{sec.3.2} above, we deduce the generalised expression of the mean and the variance of the output photon number for a given mode $j$:
\begin{equation}\label{eq51}
\langle N_{j}\rangle _{\mathrm{out}} = |\mu_{j1}|^{2}|\alpha_{1}|^{2} + \sum_{k=1}^{n} |\mu_{jk}|^{2} \sigma_{jk}  \, ,
\end{equation}
\begin{multline}\label{eq52}
\langle \Delta N_{j}^{2}\rangle _{\mathrm{out}} =  |\mu_{j1}|^{4}|\alpha_{1}|^{2} + \sum_{k=2}^{n} |\mu_{j1}\mu_{jk}|^{2} |\alpha_{1}|^{2} \\ 
+ \sum_{k=1}^{n}\sum_{l=k+1}^n |\mu_{jk}  \mu_{jl}|^{2} \sigma_{kl}\ ,
\end{multline}
where $j=1$ refers to the signal mode, $j=2,..,n$ refers to the other generated modes. The number  $\sigma_{jk}$ is equal to $0$ when the operators labeled by the integers $j$ and $k$ in Eq.\,(\ref{eq48}) are of the same type (both creation or both annihilation operators), or  to $1$ when they are of opposite types (one creation operator and one annihilation operator).


From Eqs.\,(\ref{eq24}) and (\ref{eq40}), we deduce the expression of the NF for the many-mode model in PIA configuration:
\begin{equation}\label{eq54} 
\mathrm{NF}_{j,\mathrm{PIA}} = 1 +  \frac{\displaystyle{\sum_{k=2}^{n}|\mu_{jk}|^{2}}}{|\mu_{j1}|^{2}} \, .
\end{equation}
This equation has already been derived in Ref.\,\cite{Colin}.  We remind it here for the sake of comparison. However, the generalisation to the case of the PSA has not been derived in Ref.\,\cite{Colin}, and is the subject of the following Subsection.

\subsubsection{Phase sensitive amplification}
We suppose that both the signal (mode 1) and idler (mode 2) are injected with a coherent state at the input of the fibre, the other modes being injected with vacuum state. The general expressions of the mean value and the variance of the output number of photons of mode $j$ are deduced from subsections \ref{sec.3.1} and \ref{sec.3.2} and generalised as:
\begin{multline}\label{eq55}
\langle N_{j}\rangle _{\mathrm{out}} = \sum_{k=1}^{n} |\mu_{jk}|^{2}\bigl( |\alpha_{k}|^{2}+\sigma_{jk}\bigr) \\
 + \sum_{\substack{k=1}}^{n}\sum_{\substack{l=k+1}}^{n} \bigl(\mu_{jk}\mu_{jl}^{*}\alpha_{k}\alpha_{l} + \mu_{jk}^{*}\mu_{jl}\alpha_{k}^{*}\alpha_{l}^{*}\bigr)\, ,
\end{multline}
\begin{multline}\label{eq56}
\langle \Delta N_{j}^{2}\rangle _{\mathrm{out}} =
\sum_{k=1}^{n} |\mu_{jk}|^{4} |\alpha_{k}|^{2}
\\ +
 \sum_{\substack{k=1}}^{n}\sum_{\substack{l=k+1}}^{n}|\mu_{jk}\mu_{jl}|^{2} \Bigl(|\alpha_{k}|^{2}+|\alpha_{l}|^{2}+\sigma_{kl}\Bigr)\\
+ \sum_{\substack{k=1}}^{n}\sum_{\substack{l=k+1}}^{n}  \Bigl(\mu_{jk}\mu_{jl}^{*}\alpha_{k}\alpha_{l} + \mu_{jk}^{*}\mu_{jl}\alpha_{k}^{*}\alpha_{l}^{*}\Bigr)\\
\times \Bigl(|\mu_{jk}|^{2} + |\mu_{jl}|^{2}
+  \sum_{m=l+1}^{n} |\mu_{jm}|^{2}\Bigr) \, .
\end{multline}
Substituting Eqs.\,(\ref{eq17}), (\ref{eq55}), and (\ref{eq56}) into Eq.\,(\ref{eq23}), and considering that the signal and idler are injected at the fibre input with the same input number of photons $|\alpha|^{2}\gg 1$, the generalised form of the PSA NF in the case of the many-mode model is given by:
\begin{multline}\label{eq57}
\mathrm{NF}_{j,\mathrm{PSA}} =
\left\{\sum_{k=1}^{2} |\mu_{jk}|^{4}
+  \sum_{\substack{k=1}}^{n}\sum_{\substack{l=k+1}}^{n} |\mu_{jk}\mu_{jl}|^{2} \bigl(y_{k} + y_{l}\bigr)\right.\\
\left.\hspace{-1cm} + 2 |\mu_{j1}\mu_{j2}|  \cos{\Theta} \Bigl(|\mu_{j1}|^{2} + |\mu_{j2}|^{2}
+  \sum_{m=3}^{n} |\mu_{jm}|^{2}\Bigr)\right\}\\
\times 
\left(\frac{1}{\displaystyle{\sum_{k=1}^{2} |\mu_{jk}|^{2} 
+ 2 |\mu_{j1}\mu_{j2}| \cos{\Theta}}} \right)^{2} \, ,
\end{multline}
where $y_{k}=1$ if mode $k$ is injected at the input, i.e., $k=1,2$, otherwise $y_{k}=0$, and $\Theta =\theta_{j1}-\theta_{j2}+\theta_{1}+\theta_{2}$ is the input relative phase between the pumps, signal, and idler.


Now, we address the more general case where we suppose that an arbitrary number $p$ of modes, including the signal, are injected with a coherent state at the input of the fibre, the rest being injected with the vacuum state. Then, the general expressions of the mean value and the variance of the output number of photons of mode $j$ are obtained using the \textit{Mathematica} code given in the supplementary information:
\begin{multline}\label{eq55a}
\langle N_{j}\rangle _{\mathrm{out}} = \sum_{k=1}^{n} |\mu_{jk}|^{2}\bigl( |\alpha_{k}|^{2}+\sigma_{jk}\bigr) \\
 + \sum_{\substack{k=1}}^{n}\sum_{\substack{l=k+1}}^{n} A_{kl}\bigl(\mu_{jk}\mu_{jl}^{*}\alpha_{k}\alpha_{l}+\mu_{jk}^{*}\mu_{jl}\alpha_{k}^{*}\alpha_{l}^{*}\bigr)\\
 + \sum_{\substack{k=1}}^{n}\sum_{\substack{l=k+1}}^{n}  B_{kl}\bigl(\mu_{jk}^{*}\mu_{jl}\alpha_{k}\alpha_{l} + \mu_{jk}\mu_{jl}^{*}\alpha_{k}^{*}\alpha_{l}^{*}\bigr) \\
 + \sum_{\substack{k=1}}^{n}\sum_{\substack{l=k+1}}^{n} C_{kl}\bigl(\mu_{jk}\mu_{jl}^{*}\alpha_{k}\alpha_{l}^{*} + \mu_{jk}^{*}\mu_{jl}\alpha_{k}^{*}\alpha_{l}\bigr)\\
 + \sum_{\substack{k=1}}^{n}\sum_{\substack{l=k+1}}^{n} D_{kl}\bigl(\mu_{jk}\mu_{jl}^{*}\alpha_{k}^{*}\alpha_{l} + \mu_{jk}^{*}\mu_{jl}\alpha_{k}\alpha_{l}^{*}\bigr) \, ,
\end{multline}
\begin{multline}\label{eq56a}
\langle \Delta N_{j}^{2}\rangle _{\mathrm{out}} =
\sum_{k=1}^{n} |\mu_{jk}|^{4} |\alpha_{k}|^{2} 
\\ +
 \sum_{\substack{k=1}}^{n}\sum_{\substack{l=k+1}}^{n}|\mu_{jk}\mu_{jl}|^{2} \Bigl(|\alpha_{k}|^{2}+|\alpha_{l}|^{2}+\sigma_{kl}\Bigr)\\
+ \Biggl\{ \sum_{\substack{k=1}}^{n}\sum_{\substack{l=k+1}}^{n} A_{kl}\bigl(\mu_{jk}\mu_{jl}^{*}\alpha_{k}\alpha_{l}+\mu_{jk}^{*}\mu_{jl}\alpha_{k}^{*}\alpha_{l}^{*}\bigr)\\
 + \sum_{\substack{k=1}}^{n}\sum_{\substack{l=k+1}}^{n} B_{kl} \bigl(\mu_{jk}^{*}\mu_{jl}\alpha_{k}\alpha_{l} + \mu_{jk}\mu_{jl}^{*}\alpha_{k}^{*}\alpha_{l}^{*}\bigr) \\
 + \sum_{\substack{k=1}}^{n}\sum_{\substack{l=k+1}}^{n} C_{kl} \bigl(\mu_{jk}\mu_{jl}^{*}\alpha_{k}\alpha_{l}^{*} + \mu_{jk}^{*}\mu_{jl}\alpha_{k}^{*}\alpha_{l}\bigr)\\
 + \sum_{\substack{k=1}}^{n}\sum_{\substack{l=k+1}}^{n} D_{kl}\bigl(\mu_{jk}\mu_{jl}^{*}\alpha_{k}^{*}\alpha_{l} + \mu_{jk}^{*}\mu_{jl}\alpha_{k}\alpha_{l}^{*}\bigr) \Biggr\} \\
\times \Bigl(|\mu_{jk}|^{2} + |\mu_{jl}|^{2}
+  \sum_{m=l+1}^{n} |\mu_{jm}|^{2}\Bigr) \, ,
\end{multline}
where we have introduced the following notations:
\begin{itemize}
    \item  $A_{kl}=1$ when the operator labeled by the integer $k$ in Eq.\,(\ref{eq48}) is an annihilation operator and the operator labeled by the integer $l$ is a creation operator, otherwise  $A_{kl} = 0$.
    \item $B_{kl}=1$ when the operator labeled by the integer $k$ is a creation operator and the operator labeled by the integer $l$ is an annihilation operator, otherwise  $B_{kl} = 0$.
    \item $C_{kl}=1$ when the operators labeled by the integers $k$ and $l$ are both annihilation operators, otherwise $C_{kl} = 0$.
    \item $D_{kl}=1$ when the operators labeled by the integers $k$ and $l$ are both creation operators, otherwise  $D_{kl} = 0$.
\end{itemize}
Substituting Eqs.\,(\ref{eq17}), (\ref{eq55a}), and (\ref{eq56a}) into Eq.\,(\ref{eq23}), and considering that only the $p$ first modes ($j=1\cdots p$), with $p\leq n$, are injected at the fibre input with the same coherent state $\ket{\alpha}$ with $|\alpha|^{2}\gg 1$, the generalised form of the PSA NF in the case of the many-mode model is given as:
\begin{multline}\label{eq57a}
\mathrm{NF}_{j,\mathrm{PSA}} =  
\left\{\sum_{k=1}^{p} |\mu_{jk}|^{4}
+ \sum_{\substack{k=1}}^{n}\sum_{\substack{l=k+1}}^{n} |\mu_{jk}\mu_{jl}|^{2} \bigl(y_{k} + y_{l}\bigr)\right.\\
\left.\hspace{-1cm} + \sum_{\substack{k=1}}^{p-1}\sum_{\substack{l=k+1}}^{p}  2 |\mu_{jk}\mu_{jl}|  \Bigl(A_{kl}\cos{\Theta_{A}} + B_{kl}\cos{\Theta_{B}} + C_{kl}\cos{\Theta_{C}} \right.\\
\left.\hspace{-1cm} D_{kl}\cos{\Theta_{D}} \Bigr) \Bigl(|\mu_{jk}|^{2} + |\mu_{jl}|^{2}
+  \sum_{m=l+1}^{n} |\mu_{jm}|^{2}\Bigr)\right\} \times \Biggl(\frac{1}{\Omega}\Biggr)^{2} \, ,
\end{multline}
where $\Omega$ is given by the following expression:
\begin{multline}
    \Omega = \sum_{k=1}^{p} |\mu_{jk}|^{2} 
+ \sum_{\substack{k=1}}^{p-1}\sum_{\substack{l=k+1}}^{p} 2|\mu_{jk}\mu_{jl}| \Bigl(A_{kl}\cos{\Theta_{A}} \\
+ B_{kl}\cos{\Theta_{B}} + C_{kl}\cos{\Theta_{C}} + D_{kl}\cos{\Theta_{D}} \Bigr) \, ,
\end{multline}
The relative phases between the pumps and the modes labeled $k$ and $l$ are expressed as: 
\begin{equation}
\left\{
    \begin{array}{ll}
        \Theta_{A} = \theta_{jk} - \theta_{jl} + \theta_{k} + \theta_{l} \\ 
     \Theta_{B} = -\theta_{jk} + \theta_{jl} + \theta_{k} + \theta_{l} \\
     \Theta_{C} = \theta_{jk} - \theta_{jl} + \theta_{k} - \theta_{l} \\
     \Theta_{D} = \theta_{jk} - \theta_{jl} - \theta_{k} + \theta_{l}
    \end{array}
\right.
\end{equation}
where $\theta_{k}$ and $\theta_{l}$ are the input phases of the modes $k$ and $l$, respectively.

Equations (\ref{eq57}) and (\ref{eq57a}) permit a direct calculation of the NF in the case of many-mode PSAs, without resort to cumbersome analytical calculations of the mean and the  variance of the photon number. Equation (\ref{eq57}) can be used when only the signal and idler are injected with coherent states at the input $(p=2)$, while Eq. (\ref{eq57a}) can be used for an arbitrary number of injected modes $(p=1,..n)$. In the case where the hypothesis that all the excited modes are injected with the coherent state $\ket{\alpha}$ is not fulfilled, one can still use Eqs. (\ref{eq55}), (\ref{eq56}), (\ref{eq55a}), and (\ref{eq56a}).

We note that equations (\ref{eq57}) and (\ref{eq57a}) can be used for different frequency schemes of parametric amplifiers, and different degeneracy cases. Our choice of a dual-pump configuration with non degenerate signal and idler 
(see Fig.\,\ref{fig1}) is guided by the fact that this scheme is the one discussed in Ref.\,\cite{Colin}. Thus, for the sake of comparison, we focus in the following on this particular scheme.

\section{Application to the four-mode amplifier and discussion}\label{sec.4}
In this section, we make use of the general expressions of the noise figure derived in the preceding section, Eqs. (\ref{eq54}) and (\ref{eq57}), to investigate the noise figure in the case of the four-mode scheme of Fig.\,\ref{fig1}(b).

As mentioned in Section (\ref{sec.2.3.2}), when fibre dispersion is neglected, an analytical solution of the four-mode model can be found consisting in linear input-output relations. Therefore, using Eqs.\,(\ref{eq12}) and (\ref{eq13}), we calculate and plot the gains and noise figures of the four interacting modes for both PIA and PSA configurations. The case of the PIA was already investigated in Ref.\,\cite{Colin}, but not the case of the PSA, which is of particular interest for applications.

\subsection{Phase insensitive amplification}
Here, we consider that the idler and the two sidebands 1 and 2 in Fig.\,\ref{fig1}(b) are not injected at the fibre input. Thus, from Eq.\,(\ref{eq12}), the gain of mode $j$ is defined as: 
\begin{equation}\label{eq58}
    G_{j,\mathrm{PIA}} = \frac{|\hat{b}_{j}|^{2}}{|\hat{a}_{1}|^{2}} = |\mu_{j1}|^{2} \, ,
\end{equation}
where $j=1$ denotes the signal mode, and $j=2\dots4$ denote the idler and sideband modes. Note that since no light is injected in the idler and sideband modes, the so-called gains $G_{j,\mathrm{PIA}}$ for $j=2\dots4$ are actually the conversion efficiencies from the signal to the considered mode. 

Using Eq.\,(\ref{eq54}), the noise figures for the signal, idler, and sideband modes are given by:
\begin{equation}\label{eq59}
\mathrm{NF}_{j,\mathrm{PIA}} = 1 + \frac{|\mu_{j2}|^{2}+|\mu_{j3}|^{2}+|\mu_{j4}|^{2}}{|\mu_{j1}|^{2}} \, .
\end{equation}
\begin{figure}
\begin{center}
  \begin{subfigure}[b]{0.41\textwidth}
    \includegraphics[width=\textwidth]{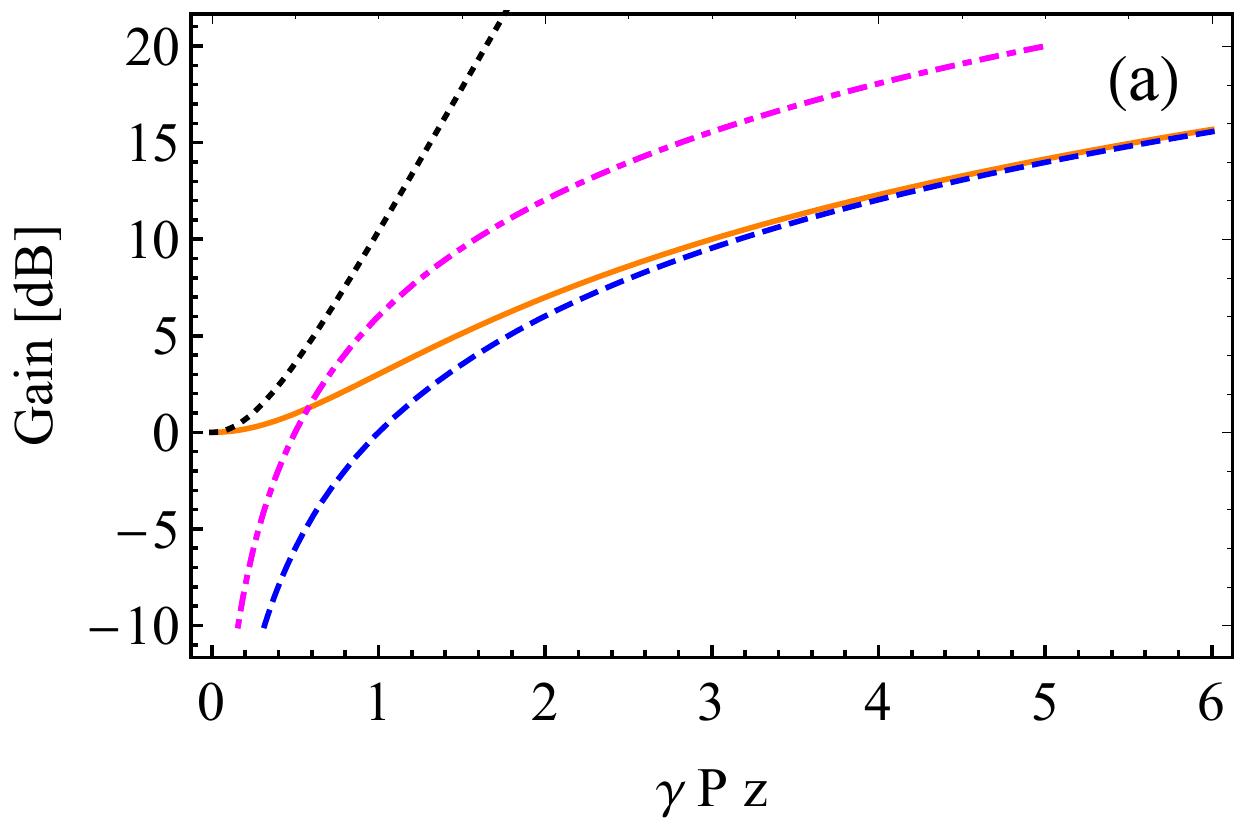}
  \end{subfigure}
  \begin{subfigure}[b]{0.4\textwidth}
    \includegraphics[width=\textwidth]{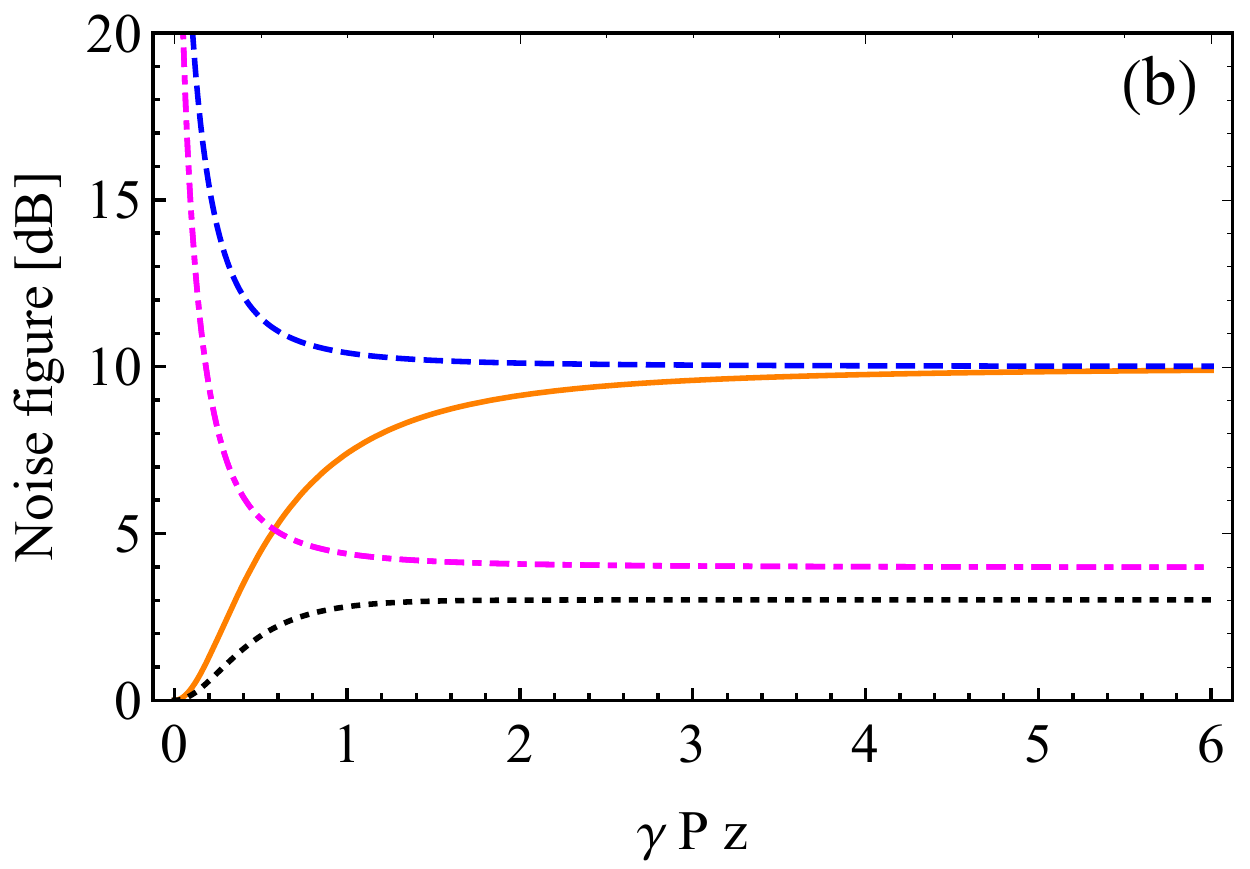}
  \end{subfigure}
  \caption{PIA configuration. Evolution of (a) the gains and (b) the noise figures versus $\gamma P z$. Dotted black line: signal in the two-mode model; Full orange line: signal in the four-mode model; Dashed blue line: sideband $2$ in the four-mode model; Dot-dashed magenta line: idler and sideband $1$ in the four-mode model.}
   \label{fig2}
\end{center}
\end{figure}
Figure \ref{fig2}(a) represents the evolution of the gains predicted by the two-mode and four-mode models as a function of the nonlinear phase $\gamma P z$. For the two-mode model, the signal gain (dotted curve) is plotted using Eq.\,(\ref{eq25}), while for the four-mode model, the gains of the four interacting modes are plotted using Eq.\,(\ref{eq58}). Since we are neglecting the fibre dispersion (see Eqs.\,(\ref{eq11}) and (\ref{eq13})), one can see that signal gains (dotted and full lines for the two-mode and four-mode models, respectively) in Fig.\,\ref{fig2}(a) grow quadratically as a function of $\gamma P z$ when the incident signal power is very small. 

Figure\,\ref{fig2}(b) reproduces the evolution of the noise figure versus $\gamma P z$, in the same situations as in Fig.\,\ref{fig2}(a). Equation\,(\ref{eq24}) is used to calculate the signal NF in the framework of the two-mode model, while the plots for the four-mode model are based on Eq.\,(\ref{eq59}). From this figure, one can see that the signal NF (dotted curve) for the two-mode case takes the well-known value of $3\,\mathrm{dB}$ when the gain becomes large for a PIA. However, the signal NF (solid curve) for the four-mode case exceeds $3\,\mathrm{dB}$ and takes a value of $10\,\mathrm{dB}$. This enhancement of the NF is due to the coupling from input vacuum noises coming from the idler and the two sidebands. Besides, a NF value equal to $10\,\mathrm{dB}$ is predicted for sideband $2$ (dashed curve), which is larger than the value of $4\,\mathrm{dB}$ predicted for the idler and sideband $1$ (dot-dashed curve). 

One notices that when $\gamma P z$ becomes large, the signal (full line) and sideband 2 (dashed line) converge to the same gains and the same noise figure value (10~dB). This is due to the strong modulation instability that occurs for the small frequency differences that we consider here between the pump and the signal \cite{Agrawal, mckinstrie2002parametric,mckinstrie2003parametric}. Indeed, as the signal propagates with the pump $1$ along the fibre, the modulation instability process produces  sideband $2$, whose power equals the signal's one for large values of $\gamma P z$ (see Fig.\,\ref{fig2}).

Besides, one notices that the mechanisms that lead to the creation of the idler and sideband 1 out of vacuum fluctuations have the same efficiency and lead to the same powers and noise figure ($4\,\mathrm{dB}$) for these two modes, as shown in Fig.\,\ref{fig2}.

\begin{figure}
\begin{center}
  \begin{subfigure}[b]{0.42\textwidth}
    \includegraphics[width=\textwidth]{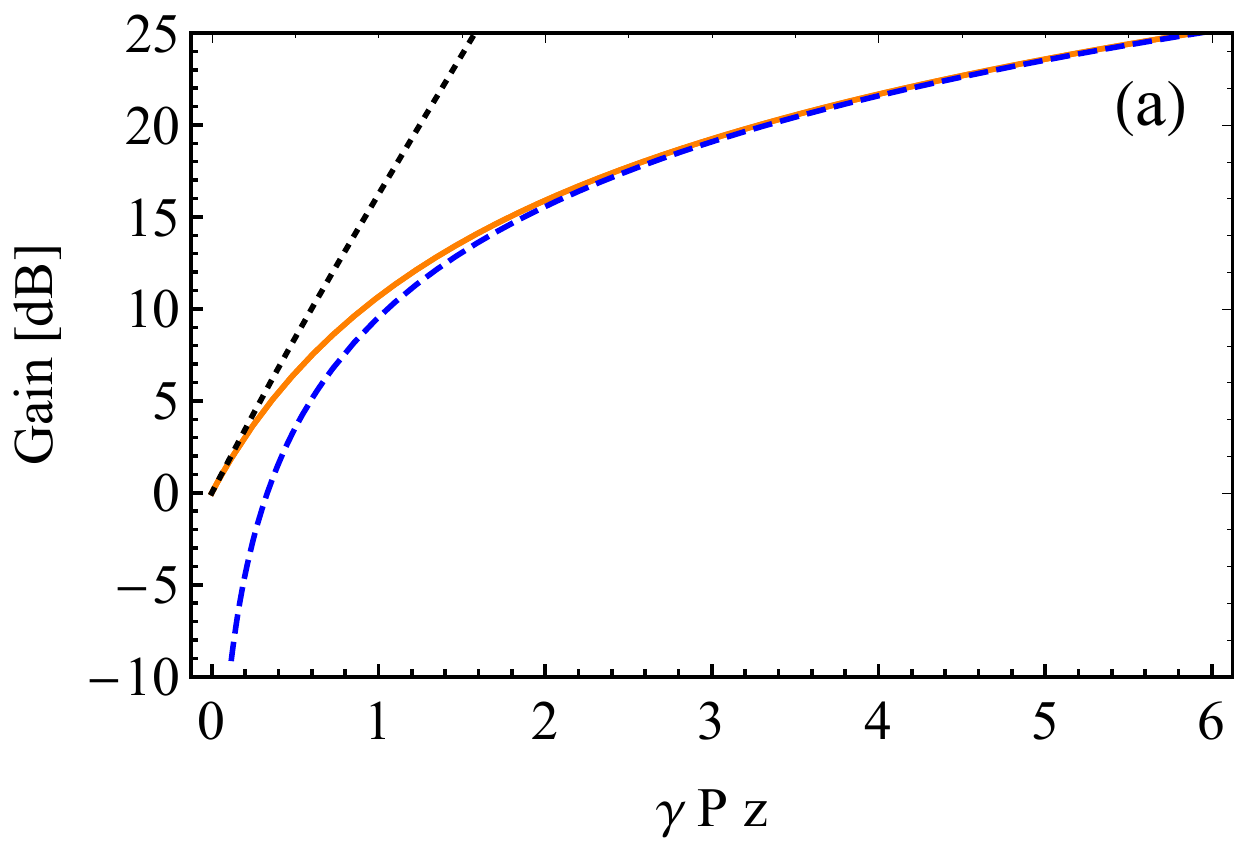}
  \end{subfigure}
  \begin{subfigure}[b]{0.41\textwidth}
    \includegraphics[width=\textwidth]{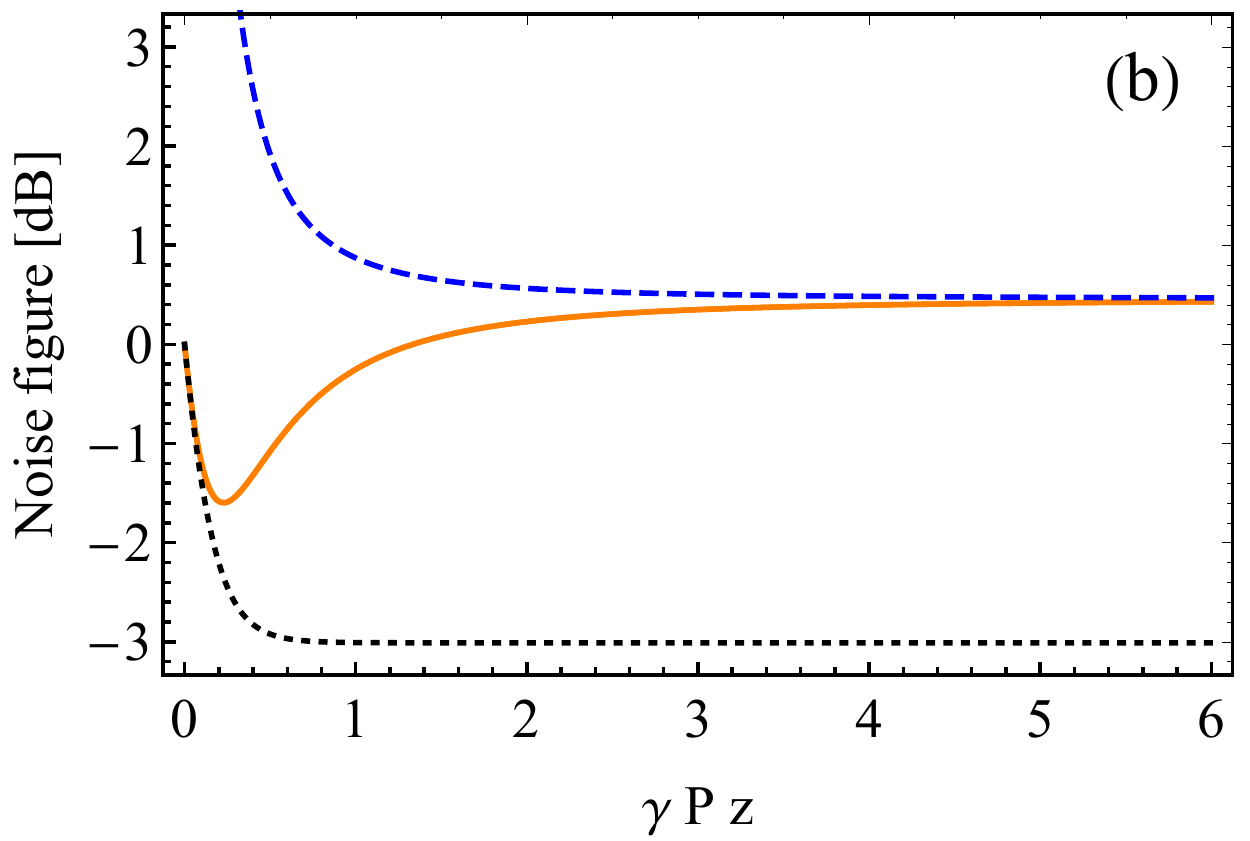}
  \end{subfigure}
  \caption{PSA configuration. Evolution of (a) the gains and (b) the noise figures versus $\gamma P z$. Dotted black line: signal and idler in the two-mode model; Full orange line: signal and idler in the four-mode model; Dashed blue line: sidebands $1$ and $2$ in the four-mode model.}
   \label{fig3} 
\end{center}
\end{figure}
\subsection{Phase sensitive amplification}
In the case of PSA operation, both the signal and idler are present at the input of the amplifier (Fig.\,\ref{fig1}(b)) and have equal amplitudes. Then, the gain of mode $j$ is given by:
\begin{equation}\label{eq60}
G_{j,\mathrm{PSA}} =  \frac{|\hat{b}_{j}|^{2}}{|\hat{a}_{1}|^{2}} = |\mu_{j1}|^{2} + |\mu_{j2}|^{2} + 2 |\mu_{j1}| |\mu_{j2}| \cos{\Theta} \, ,
\end{equation}
where $j=1$, 2, 3, and 4 denote the signal,  idler, sideband $1$, and sideband $2$, respectively.

By using Eq.\,(\ref{eq57}) for $p=2$, the noise figures of the four interacting modes are given by the following expression:
\begin{multline} \label{eq61}
\mathrm{NF}_{j,\mathrm{PSA}} = \Biggr\{|\mu_{j1}|^{4}+|\mu_{j2}|^{4}+ 2 |\mu_{j1}\mu_{j2}|^{2}  \\
+ |\mu_{j1}\mu_{j3}|^{2}  + |\mu_{j1}\mu_{j4}|^{2}  +|\mu_{j2}\mu_{j3}|^{2}  +|\mu_{j2}\mu_{j4}|^{2} \\
+ 2 |\mu_{j1}\mu_{j2}| \Bigl(\sum_{k=1,..4}|\mu_{jk}|^{2}\Bigr) \cos{\Theta}   \Biggl\}
\\
\times \frac{1}{\bigl(|\mu_{j1}|^{2} + |\mu_{j2}|^{2} + 2 |\mu_{j1}\mu_{j2}| \cos{\Theta}\bigr)^{2}} \, .
\end{multline}

Figure \ref{fig3} shows the evolutions of the gains (Fig. \ref{fig3}(a)) and noise figures (Fig. \ref{fig3}(b)) versus $\gamma P z$ for the two-mode and four-mode models in the case of PSA operation. The gains are computed using Equations (\ref{eq32}) and (\ref{eq60}) for the two-mode and four-mode models, respectively, while the noise figures are based in Equations (\ref{eq31}) and (\ref{eq61}), respectively. All calculations are performed with the value $\Theta=0$ of the relative phase between the four waves at the input of the fibre.

Contrary to Fig.\,\ref{fig2}, when the fibre dispersion is neglected, the models predict a linear increase of the  gain for small values of $\gamma P z$.

Concerning the noise figure results, Fig.\,\ref{fig3}(b) shows that the  signal and idler NF in the two-mode model (dotted curve) takes a value of -3 dB when $\gamma P z$ is large. This well known results is explained by the fact that only one of these waves is supposed to be detected. If the signal and idler  were jointly detected, the input SNR would be increased by 3 dB and thus the NF would reach the well-known quantum limited value of 0 dB.

The four-mode model predicts a degradation of the signal and idler noise figures [see the full line in Fig.\,\ref{fig3}(b)] with respect to the two-mode model. The discrepancy between the two models occurs when the nonlinearity is large enough (typically $\gamma P z\geq 0.2$) for the powers of the sidebands to be no longer negligible with respect to the signal and idler powers (see Fig.\,\ref{fig3}(a)). This degradation, which reaches 3.4~dB for large values of $\gamma P z$, is due to a transfer of the vacuum fluctuations that are injected in the sideband modes, which becomes efficient when the modulation instability creates significant sidebands.

\begin{figure}
\begin{center}
  \begin{subfigure}[b]{0.47\textwidth}
    \includegraphics[width=\textwidth]{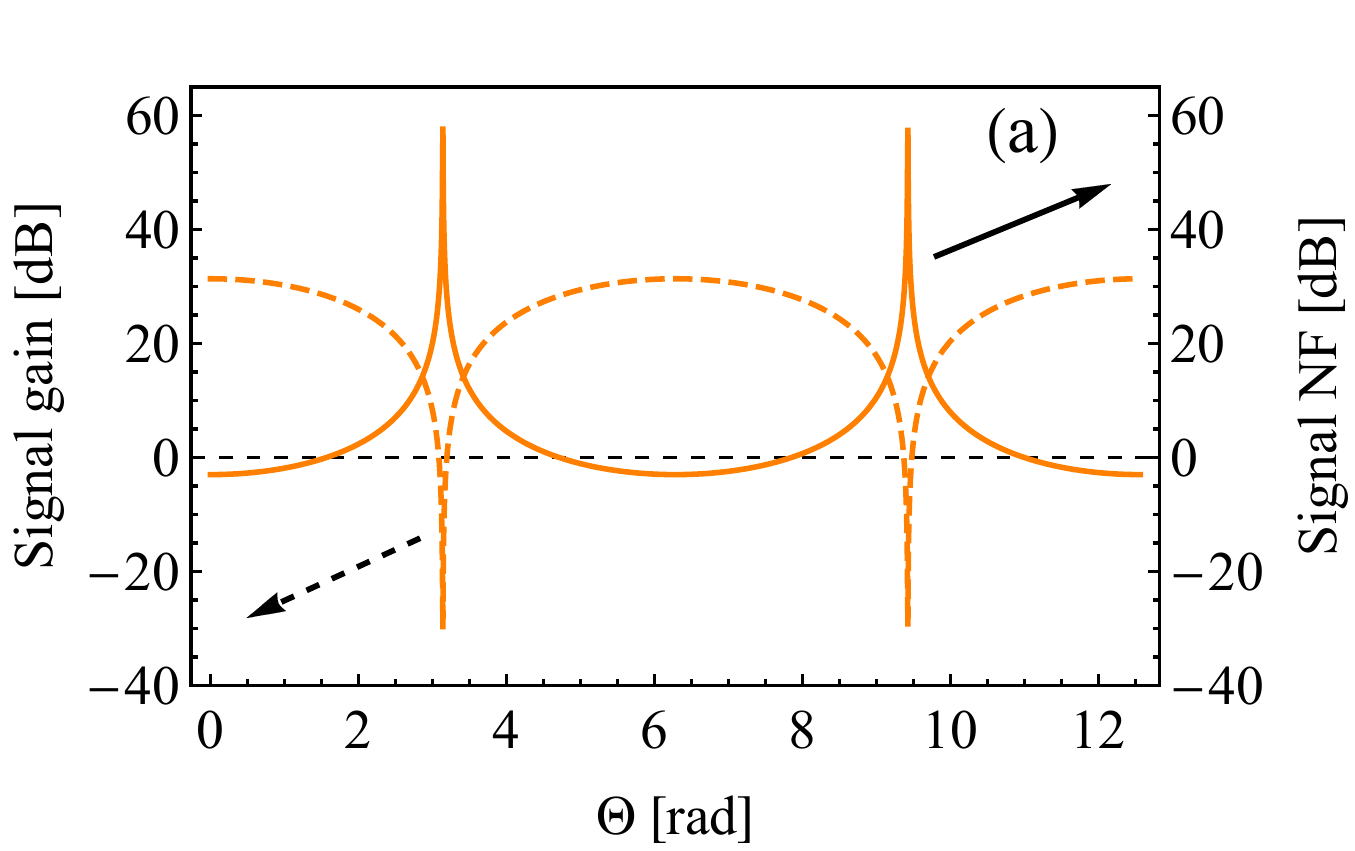}
  \end{subfigure}
  \begin{subfigure}[b]{0.45\textwidth}
    \includegraphics[width=\textwidth]{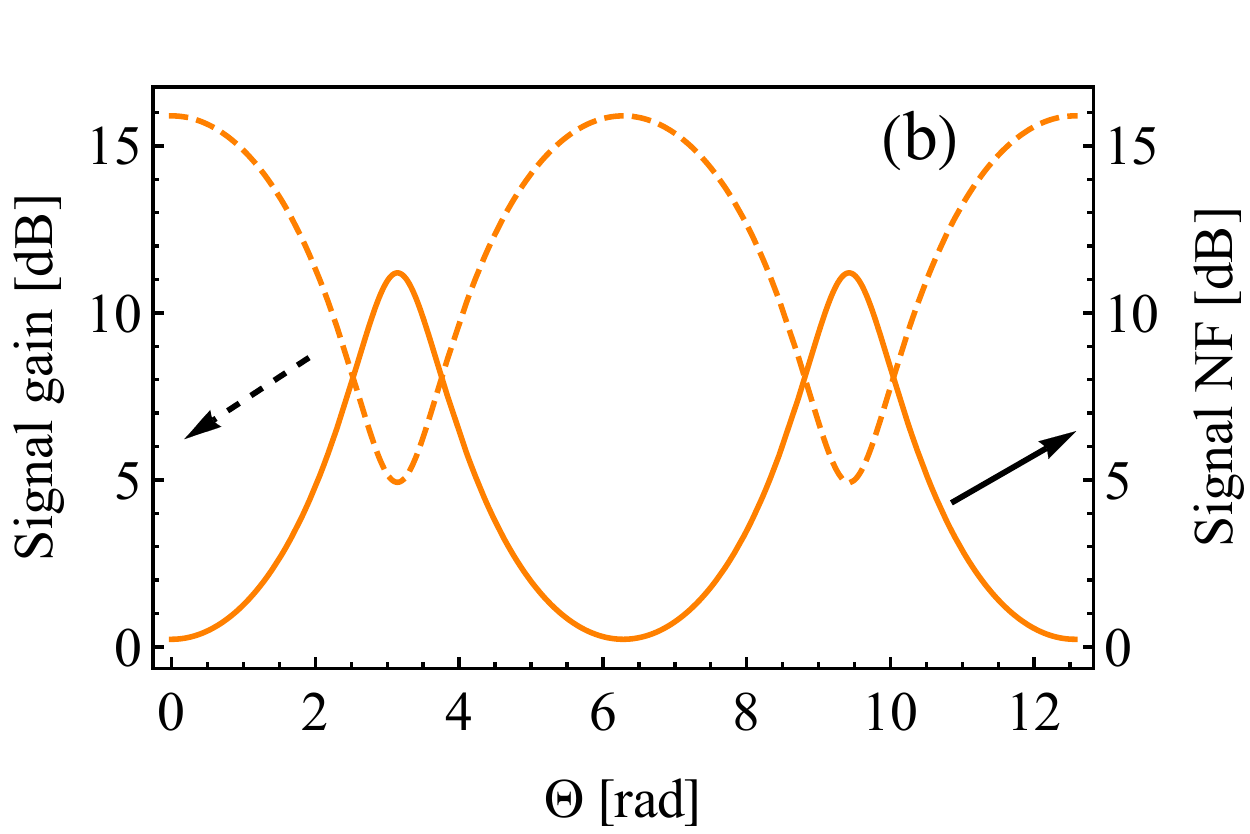}
  \end{subfigure}
  \caption{Signal gain (dashed line) and signal NF (solid line) versus relative phase $\Theta$ for (a) the two-mode model, (b) the four-mode model, ($\gamma P z =2$).}
\label{fig4}
\end{center}
\end{figure}

For such a relatively large value of $\gamma P z$,  Fig. \ref{fig4} shows the evolution versus the relative phase $\Theta$ of the signal gain and NF in the case of the two-mode [Fig. \ref{fig4}(a)] and four-mode [Fig. \ref{fig4}(b)] models for the PSA. Figure  \ref{fig4}(b) shows that the presence of the two sidebands, leads to the fact that the PSA gain is always larger than 0 dB for all values of the phase $\Theta$. This leads in particular to the fact that the minimum gain is no longer the inverse of the maximum gain. This is similar to the features already observed in the framework of a seven-wave model \cite{Weilin2017}. This phenomenon is also accompanied by a strong degradation of the maximum gain obtained for $\Theta= 2 k \pi$, where $k$ is an integer.

Concerning the noise figure, the fact that the minimum gain remains positive in the case of the four-mode model leads to the fact that maximum value of the NF is not as strong as in the case of the two-mode model. Moreover, as already seen in Fig.\,\ref{fig3}, the degradation of the minimum noise figure associated with the maximum gain is significant but not dramatic, leading to a NF equal to 0.4\,dB.


\section{Summary}\label{sec.5}
In summary, we derived generalised analytical expressions of the noise figure for many-mode phase insensitive and sensitive amplifiers using a quantum approach. The method of derivation was based on the existence of a linear relation between the input and output mode operators, which served as a starting point. We then started from some particular models involving a limited number of modes, the linear input-output relation of which could be found under certain assumptions, to derive an expression of the noise figure. We then extrapolated from these results to deduce a more general expression for the NF in the case of an arbitrary number of modes. 

As an application of this generalised NF expression, we predicted the noise figure in the case of the four-mode model for both PIA and PSA schemes. We found that the signal exhibits a higher noise level than the one expected from the conventional two-mode model, due to the vacuum fluctuations coupled to the signal from the high-order modes. In spite of this degradation of the noise figure, we could predict that a PSA in multi-mode configuration can still exhibit relatively low NF values, and in particular much lower than in PIA configuration.

Beyond applications to telecom and microwave photonics systems, the present work opens interesting perspectives to low-noise amplification of frequency combs \cite{slavik2012processing} and low-noise spectral replication \cite{richter2012experimental, tong2012broadband}, for many applications such as  metrology, THz regeneration, signal coherent communications, etc.,  where low noise levels are mandatory.




\section*{Acknowledgments}
Y. Hassouni would like to acknowledge the partial support from ICTP through AF-14.



\bibliography{cas-refs}

\end{document}